\documentclass{aastex631}
\usepackage{multirow} 

\newcommand\hd{{HD\,317101A} }   %placeholder name
\newcommand\hdp{{HD\,317101A}}   %placeholder name
\newcommand\radio{{J180526$-$292953} }

\begin{document}

\title{A Radio Flaring, Chromospherically-Inactive K Dwarf}

\author[0009-0008-7604-003X]{Dale A. Frail}
\affiliation{National Radio Astronomy Observatory, P.O. Box O, Socorro, NM 87801, USA}

\author[0009-0006-5070-6329]{Scott D. Hyman}
\affiliation{Department of Engineering and Physics, Sweet Briar College, Sweet Briar, VA 24595, USA}

\author[0000-0003-2565-7909]{Michele L. Silverstein}
\altaffiliation{NRC Research Associate}
\affiliation{U.S.\ Naval Research Laboratory,  4555 Overlook Ave SW,  Washington,  DC 20375,  USA}

\author[0000-0003-3272-9237]{Emil Polisensky}
\affiliation{U.S.\ Naval Research Laboratory,  4555 Overlook Ave SW,  Washington,  DC 20375,  USA}

\author[0000-0002-4039-6703]{Evangelia Tremou}
\affiliation{National Radio Astronomy Observatory, P.O. Box O, Socorro, NM 87801, USA}

\author[0000-0002-1634-9886]{Simona Giacintucci}
\affiliation{U.S.\ Naval Research Laboratory,  4555 Overlook Ave SW,  Washington,  DC 20375,  USA}

\author[0000-0002-9061-2865]{Hodari-Sadiki Hubbard-James}
\affiliation{Agnes Scott College, 141 E. College Ave., Decatur, GA 30030}

\author[0009-0006-7309-1005]{Jacinda Byam}
\affiliation{Agnes Scott College, 141 E. College Ave., Decatur, GA 30030}

\author[0000-0002-2532-2853]{Steve~B.~Howell}
\affiliation{NASA Ames Research Center, Moffett Field, CA 94035, USA}

\author[0000-0002-4235-6369]{Robert F. Wilson}
\affiliation{NASA, Goddard Space Flight Center, Mail Code 660
Greenbelt, MD 20771}

\author[0009-0009-4693-7783]{Matthew Lastovka}
\affiliation{Department of Astronomy, University of Maryland
4296 Stadium Dr., College Park, MD 20742-2421}

\author[0000-0001-6812-7938]{Tracy E. Clarke}
\affiliation{U.S.\ Naval Research Laboratory,  4555 Overlook Ave SW,  Washington,  DC 20375,  USA}

\author[0000-0001-8035-4906]{Namir E. Kassim}
\affiliation{U.S.\ Naval Research Laboratory,  4555 Overlook Ave SW,  Washington,  DC 20375,  USA}

\correspondingauthor{Dale A. Frail}
\email{dfrail@nrao.edu}

\begin{abstract}
%247 words of the 250 word limit
We report on an unusual radio source (J180526$-$292953), initially identified as a steep spectrum, polarized point source toward the Galactic bulge and found to coincide with the nearby K dwarf \hdp. We conducted a multi-wavelength radio study utilizing new GMRT observations and archival data from ASKAP, MeerKAT, and the VLA. At 1.5 GHz, \hd exhibits highly polarized coherent emission with variable activity lasting several hours with an apparent period of 3.7 days, which is consistent with electron cyclotron maser (ECM) emission. The behavior at 3 GHz is distinctive, with a short burst lasting tens of seconds to minutes, a flat spectrum, and no detected polarization, possibly suggesting gyro-synchrotron emission. High-resolution optical spectroscopy from CHIRON/SMARTS confirms \hd as a mature, chromospherically inactive K7V star, while {\it Gaia} astrometry, combined with speckle imaging from Zorro/Gemini-S, indicates the presence of a close-in M5.5V companion. We evaluated three possible origins for the combined radio behavior: chromospheric activity, auroral emission (possibly from a star-planet interaction), or an ultra-long-period transient. The bulk of the evidence favors an auroral origin, but the dominant stellar source of the ECM emission remains uncertain. Future VLBI observations, long-term TESS monitoring, high resolution spectroscopy and further radio characterization will be key to distinguishing between various scenarios.

\end{abstract}




\section{Introduction}\label{sec:intro}

The first confirmed extrasolar planets were found using radio timing methods around degenerate stars \citep{1992Natur.355..145W,1993ApJ...412L..33T,1993Natur.365..817B}. More recently, there has been a resurgence of interest in using radio techniques to detect exoplanets around late-type stars and ultracool dwarfs (UCD) using plasma emission processes to identify magnetoionic star-planet interactions (SPI). The larger goal of such detections is to characterize the strength and topology of exoplanet magnetic fields, and to better understand how space weather (e.g., coronal mass ejections) might influence planetary atmospheres and habitability. For an up-to-date and comprehensive review of this topic see \citet{2024NatAs...8.1359C}.

The most efficient mechanism for tracing star-planet interactions is the electron cyclotron maser instability (ECM). In ECM a small fraction of the energy of high-velocity electrons spiraling in a magnetic field is converted into radio emission \citep{1979ApJ...230..621W,1982ApJ...259..844M}. The signature of ECM is highly circularly polarized emission (up to 100\%) with a high brightness temperature (T$_b>10^{12}$ K). Additionally, the radiation from ECM is emitted at close to the first or second harmonics of the gyro-frequency ($\nu_c$),
%it is expected to have a bandwidth of order unity ($\delta\nu/\nu\simeq{1}$), 
with a sharp spectral cutoff that is proportional to the magnetic polar field strength \citep{2002ARA&A..40..217G}:
\begin{equation}\label{eqn:ecm}
    \nu_c = {e \,{\rm B}\over{2\pi m_e c}} \simeq 2.8\, {\rm B}
\end{equation}
where if $\nu_c$ is given in MHz, B is the local magnetic field strength in Gauss, while $c$, $e$ and $m_e$ are the speed of light and the charge and mass of the electron, respectively. Auroral radio emission has been detected from the five magnetized planets in our solar system and interpreted as originating from ECM \citep{1998JGR...10320159Z}. In addition to the high brightness temperatures and circular polarization expected from ECM, the time-averaged spectra show a sharp drop off of the emission (with spectral slope $\alpha\simeq-5$, where S$_\nu\propto\nu^\alpha$) at frequencies above 10 kHz to 10 MHz, directly measuring the planetary B fields from Eqn.\,\ref{eqn:ecm}. Another important observational signature relevant to SPI is that the planetary ECM emission is beamed at relatively large angles to the B field (up to 90$^\circ$) and shows rotational and/or orbital periodicity. With current sensitivities, the most likely detection of SPI at radio wavelengths is not from the interaction of the stellar wind with the planetary magnetosphere \citep[analogous to the the auroral planetary emissions in our solar system;][]{2007P&SS...55..598Z}. Instead, the expected SPI signature is analogous to the Jupiter-Io system, in which a near-in planet undergoes a sub-Alfv\'enic interaction with the star's magnetic field inducing stellar auroral emission \citep{2024NatAs...8.1359C}.
% How do you want to introduce the breakdown of co-rotation "or reconnection at the edge of the stellar magnetosphere."

While a definitive detection of ECM from a magnetized exoplanet has yet to be made \citep[see][and references therein]{2024arXiv240412348L}, there are numerous examples of analogs of solar system magnetopheres among red dwarf and ultra-cold dwarf stars. For example, \citet{2015Natur.523..568H} argued for an auroral origin for the ECM emission from the M8.5 UCD LSR\,J1835+3259 and \citet{2023Natur.619..272K} have directly imaged its radiation belt. There is a consensus that near the end of the main sequence, around a spectral type of M4 where stars become fully convective, the activity makes a transition from chromospheric to auroral \citep{2017ApJ...846...75P,2024A&A...684A...3Y}. In all of the cases above it is not demonstrated that exoplanets are the magnetoplasmic engine powering this emission. 

Additionally, one of the challenges to confirming the radio detection of exoplanets is that ECM emission is not unique to SPI. ECM emission has also been seen from the stable magnetospheres of hot magnetized early-type stars \citep[e.g.,][]{2018MNRAS.474L..61D}, in chromospherically active late-type stars \citep{2019ApJ...871..214V,2019MNRAS.488..559Z,2022ApJ...935...99B}, and possibly in magnetized cataclysmic variables \citep{2020AdSpR..66.1226B,2024arXiv241024157R}. More recently, auroral-like radio emission has been detected above the sunspots on our Sun, raising the possibility of a new source of ECM emission for late-type stars with large persistent sunspots \citep{2024NatAs...8...50Y}. Among dwarf stars, alternatives such as the breakdown of co-rotation or magnetic re-reconnection high up in the stellar magnetopshere may be to blame \citep{2024NatAs...8.1359C}.

Several innovative approaches have been taken to overcome these limitations and bolster SPI claims. One method has been to use indicators of stellar activity (e.g., H$\alpha$ emission, Ca H and K lines, rapid rotation, etc.) to flag chromospherically active stars that are likely to exhibit ECM from flares. Samples have been formed of quiescent red dwarfs, including the SPI candidate GJ\,1151 \citep{2020NatAs...4..577V,2021NatAs...5.1233C,2021ApJ...919L..10P}. A second promising method uses the fact that the auroral emission from solar system planets is beamed and to search for {\it periodic} ECM. Since long-lived sunspots can show periodicities, it is useful if the orbital period of the planet(s) are known before hand, despite the fact that the orbital phase during which the beamed emission appears can be a complex function of the orbital geometry, and rotation/orbital periods of the star and planet(s)
 \citep{2023MNRAS.524.6267K}. While periodic ECM emission has been seen from several M dwarfs \citep{2019MNRAS.488..559Z,2024MNRAS.531..919Z,2024A&A...682A.170B}, only the M4.5 star YZ Ceti has shown polarized, coherent bursts consistent with the period of its innermost planet \citep{2023NatAs...7..569P}. A third possible discriminant that could distinguish ECM emission from stellar flares (and sunspots) from SPI is to look for differences in their spectral and polarization properties. The coronal shocks that give rise to flares are fundamentally stochastic, sampling different plasma and magnetic field configurations from one flare to the next. In contrast, in a relatively stable polar magnetosphere (such as that which might be expected in SPI) we might expect a persistence in the time-averaged spectral and polarization properties of the ECM emission \citep{2019ApJ...871..214V,2021A&A...648A..13C}. While the  dynamic spectra of the auroral radio emission in our solar system show significant structure (owing to viewing angle changes in the beamed emission cone), the properties of the time-averaged EM emission are relatively stable \citep{1998JGR...10320159Z}.

% Need to read about solar systemorigins of dynamic spectra. sufficient but not necessary as the B field geometry could be complex

In this paper we report on the anomalous K dwarf \hd nicknamed ``Special K''. It was initially identified as J180526$-$292953 in a polarimetric radio imaging survey of the Galactic bulge area, and was flagged as an outlier for its unusually high circular polarization and steep spectral index \citep{2024ApJ...975...34F}.  Despite preliminary indications that \hd was a normal quiescent main sequence dwarf star, it was unusually radio loud. The peculiar radio properties of \hd motivated us to undertake a more detail multi-wavelength study of this system. In \S\ref{sec:obs} we carry out a more detailed analysis of these discovery data, as well as describing new and archival observations at radio, optical and X-ray wavelengths. We discuss models for their origins in \S\ref{sec:discuss}, while in  \S\ref{sec:conclude} we summarize the strengths and weaknesses of the proposed models, and suggest possible future directions.

\section{Observations}\label{sec:obs}

\subsection{Radio Observations}\label{sec:radio}

An extensive series of new and archival observations were analyzed in order to study the radio emission from J180526$-$292953. A summary of these observations are given in Table \ref{tab:radioobs}. We list the telescope used, the program name, the observing dates or range, the number of epochs, the frequency range, and the time-on source (ToS) per pointing and the angular resolution ($\theta$). A full description of the discovery of this stellar radio source and its properties are given below in \S\ref{sec:meerkat}-\S\ref{sec:archive}.

\begin{table}[!ht]
    \caption{Summary of Radio Observations Taken Toward \hdp}\label{tab:radioobs}
    \centering
    \begin{tabular}{lcrccrcr}
        \hline
        Telescope & Program & Date~~~~~~~~~ & Epochs & Freq. & ToS & $\theta$  \\
        \omit & \omit & (YYYY-MM-DD) & \omit & (GHz) & (min) & (arcsec) \\
        \hline
        \hline
        MeerKAT & SSV-20180505-FC-01 & 2019-2020 & 4 & 0.9-1.7 & 54 & 7\\
        MeerKAT & ThunderKAT & 2021 & 30 & 0.9-1.7 & 12-18 & 7 \\
        GMRT & DDT C352 & 2024-06-02 & 1 & 0.55-0.90 & 30 & $3.1\times7.5$ \\
        GMRT & DDT C352 & 2024-06, 07 & 4 & 0.95-1.46 & 30 & 2$\times{5}$ \\
        GMRT & TGSS & 2010-04-28 & 1 & 0.14-0.16 & 100 & 25\\
        VLA & VLASS & 2016-23 &  6 & 2.0-4.0 & 0.2 & 2\\
        VLA & VLITE & 2023-06-14 & 1 & 0.32-0.36 & 0.5 & 20 \\
        VLA & NVSS & 1996-05-10 & 1 & 1.36-1.45 & 1 & 45 \\ 
        ASKAP & VAST-Low & 2019-24 &  50 & 0.7-1.0 & 12 & 15\\
        ASKAP & VAST-Mid & 2021 & 3 & 1.2-1.5 & 12 & 8 \\
        ASKAP & RACS-Low & 2019, 2024 & 3 & 0.7-1.0 & 15 & 25 \\
        ASKAP & RACS-Mid & 2021-07-29 & 1 & 1.3-1.5 & 15 & 10\\
        ASKAP & RACS-High & 2022-01-23 & 1 & 1.5-1.7 & 15 & 7\\
%        MOST & SUMSS & 2002 & 1 & 0.82-0.85 & 720 & 45 \\
        \hline
    \end{tabular}
    \vspace{1cm}
%    \tablecomments{As VLASS epochs were made in scan mode, see text for time-on-source.}
\end{table}

\subsubsection{MeerKAT Archival Observations}\label{sec:meerkat}

There were two datasets taken with the MeerKAT Radio Telescope Array. The discovery data come from a 856--1712 MHz (L-band) survey of the bulge of our Galaxy (Code SSV-20180505-FC-01) carried out during 4 epochs
%42 sessions 
from 2019 December 26 to 2020 August 05 (mean epoch 2020.3) \citep{2025ApJ...985...94C}. An additional 30 sessions of data were taken from 856--1712 MHz as part of the ThunderKAT Large Science Project from 2021 May 04 to 2021 December 18 \citep{2016mks..confE..13F}.

The radio source \radio was discovered in two overlapping Galaxy bulge mosaic 3.125$^\circ$×3.125$^\circ$ images \citep{2024ApJ...975...34F}. The method of observation and the data reduction are detailed in \citet{2024ApJ...975...34F} and \citet{2025ApJ...985...94C}. Its position in Galactic coordinates is ($l, b$)=(1.69$^\circ$, $-4.02^\circ$), or in J2000 coordinates R.A.=18$^h$ 05$^m$ 26.44$^s$, Dec.= $-$29$^\circ$ 29$^{\prime}$ 54.6$^{\prime\prime}$ (mean epoch 2020.3). The position uncertainty is approximately $0.7^{\prime\prime}$, which includes a measurement error added in quadrature with a systematic astrometric error.
After correcting for the proper motions of all {\it Gaia} stars in the vicinity, we found that the MeerKAT radio source coincided with the star \hd \citep[see][for more details]{2024ApJ...975...34F}.

%The {\it mean} flux density (I), spectral index ($\alpha$) and circular polarization (V/I) are {\bf 0.17$\pm$0.01 mJy, $-$4.28$\pm$0.38, +89.7\% (S/N=20)} \sout{0.15$\pm$0.01 mJy, $-$4.25$\pm$0.38, +93.1\% (S/N=19.1) and 0.18$\pm$0.01 mJy, $-$4.31$\pm$0.38, and +86.3\% (S/N=21.5)}, {\bf for} \sout{in} two {\bf overlapping} \sout{adjacent} mosaic images\sout{, respectively}. No linear polarization was detected with $3\sigma$ limits of 7.5\%.
%I did not reduce the uncertainties by sqrt(2) since the noise in the overlapping region should be correlated.

\begin{figure}[!ht]
\centering
\includegraphics[width=0.5\textwidth,angle=0,trim=0cm 0cm 0cm 0cm]{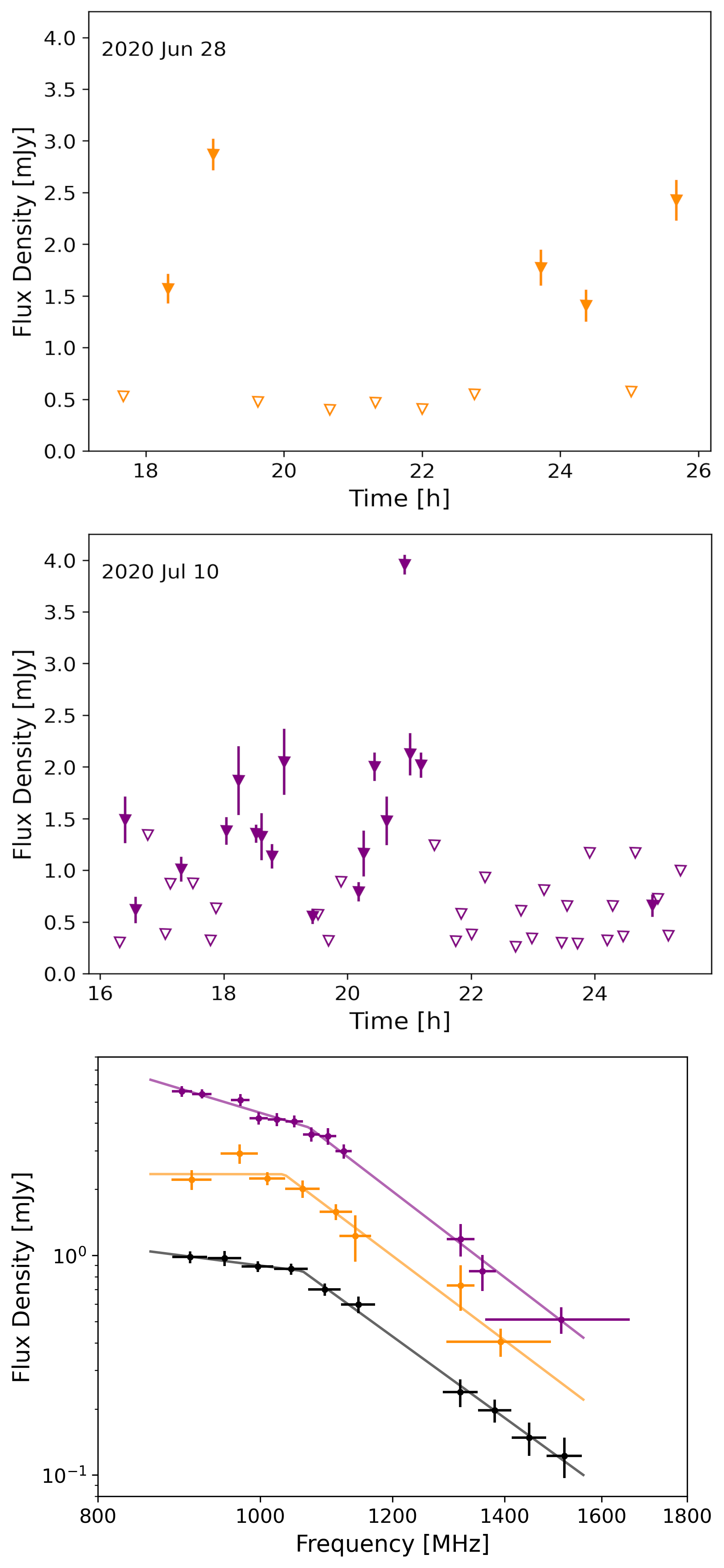}
\caption{Top: Stokes RR light curve from 2020 June 28 using the lower half of the band centered at 1.022 GHz. Detections are shown as filled triangles; 3-$\sigma$ upper limits are indicated by open triangles. The observing session spans 9 hours and consists of a single pointing with twelve 4.5-minute scans. Middle: Same as top, but for 2020 July 10. This 9 hour observing session includes four pointings, totaling 48 scans of 4.5 minutes each. Bottom: Average Stokes RR spectra with broken power-law model fits from 856 MHz to 1712 MHz. The orange (middle) curve corresponds to all scans with detections from June, the purple (top) curve to the brightest 5-minute scan from July, and the black (bottom) curve from the average of all one-hour pointings.}
\label{fig:meerkat}
\vspace{1cm}
\end{figure}

\hd varies on month-long timescales. 
%The mosaic images containing the radio source were constructed from several MeerKAT pointings{\bf, each with a $\sim1.0\deg$ FWHM primary beam, and consisting of 12 x 4.5-min scans observed over 9 hours. 
The source is detected on 2020 June 28 and July 10, and is not detected on 2019 December 31 and 2020 January 04.
%three are at or within the $\sim0.5\deg$ half-power point of the primary beam on 2020 June 28 and July 10, and a fourth at $\sim0.8\deg$ also on July 10. 
%\sout{The source is detected on 2020 June 28 and July 10 in three pointings at or within the $\sim0.5\deg$ half-power point of the primary beam, and three others at $0.8-1.1\deg$.}
%\sout{Each consists of 12 x 4.5-min scans observed over 9 hours. Two additional observations on 2019 Dec.\,31 and 2020 Jan.\,04, pointed $\sim1\deg$ away, do not detect the source.} 
The mean flux density (or limits) for each of these 4 epochs is listed in Table \ref{tab:radio}.
% DF: I don't think there is a 50% variation between June and July epochs based on the table. SDH: Imaging RR in the lowest IFs shows June 28 is about 25% higher than July 10. Difficult to show that for the average Stokes I over the whole band. But in the end, it's the light curves that show the comparison. So no need for this sentence, although I think we could leave it in the already submitted Mosaic-based paper, changed to 25%. 
A search for shorter time scale variability within the 2020 June 28 and July 10
%\sout{two observing} 
epochs
%\sout{when the source was detected (2020 June 28 and July 10) were} 
{was} undertaken. We found that imaging the lower half of the band (center frequency 1.022 GHz) and Stokes RR led to a significant increase in signal-to-noise owing to the extremely high degree of polarization and very steep spectrum. Over each of these 9 hr intervals there are periods of $\sim$4 hr duration when \hd is active, with significant fluctuations on hour-to-hour timescales. Scan-to-scan fluctuations of 1 to 4 mJy, with many non-detections interspersed, are shown in the 2020 June 28 and July 10 light curves in the upper panels of {Fig.\,\ref{fig:meerkat}}. 
%\sout{(Scans for the detection in the June 28 pointing located $\sim1\deg$ away are not plotted since the source is only marginally detected in a few individual scans.)} 

We note the sharp decrease from the 4 mJy peak at 20.929 hr (UTC) on 2020 July 10 to 2 mJy in the next scan beginning only 30-sec later. We looked for but did not find significant flux variations on intra-scan ($<5$ min) timescales during the bright 2 hr long ``flare'' on July 10. If we exclude the time intervals on both 2020 June 28 and July 10 when the source is clearly detected, and we image the combination of scans for which we have only upper limits, the source is still detected at $\sim0.3$ mJy (5$\sigma$). This suggests that the radio emission is present at low levels throughout each 9-hr observation. 

One of the more remarkable properties of the radio source is that its spectral and polarization properties remain similar at different epochs and at different activity levels. We discuss this in more detail below, but within the errors the spectral index and both the degree and handedness of the circular polarization are the same on 2020 June 28 and July 10, with no significant change on any of the timescales that we sampled. In the bottom panel of Fig.\,\ref{fig:meerkat} we show the mean spectrum obtained from all the pointings (bottom, black) along with a spectrum formed from the average of the 5 June detections (middle, orange) and a spectrum formed from the brightest 5 minutes of data on 2020 July 10 (top, purple).

The mean spectrum can be fit by a simple power law from 856 MHz to 1712 MHz, S$_\nu$=S$_\circ\,\nu^{-4.4}$ mJy, but there is a suggested turnover at the low end of the band. If we fit a piecewise broken power law to these data the quality of the fits improves with a reduced $\chi^2$ for the June, July and average spectrum (bottom panel of Fig.\,\ref{fig:meerkat}.
The power law slope of the low end of the spectrum is not well constrained, but above the break it is $-5.74\pm 0.57$, $-5.84\pm 0.39$ and $-5.53\pm 0.39$, respectively. The corresponding break frequencies ($\nu_c$) are 1034$\pm$20 MHz, 1070$\pm$21 MHz, and 1060$\pm$19 MHz. Regardless of whether we use all the pointing data, the June detections, or just the 5 minutes of peak activity in July, the spectrum remains the same within the errors. We also attempted to detect short-term spectral variability by further subdividing the data in time and frequency, but we lacked the S/N to make a dynamic spectrum.

%\sout{of 1.38, 0.55 and 0.22, respectively. 

{We derive the degree of polarization for each epoch from the lower half of the band where there is signal.  The values of V/I, if we include {\it all} the nearest pointing data in Fig.\,\ref{fig:meerkat}, is +94$\pm$13\% and +78$\pm$8\% for 2020 June 28 and 2020 July 10, respectively.
%\sout{We can improve the signal to noise of these V/I values if we just include the data for when the source is brightest. Combining just the five detections on June 28 yields +95.7$\pm$7.5\%. Combining all detected scans of the nearest pointing on July 10 yields +74.8$\pm$7.5\%, while the} 
The brightest scan alone on July 10 yields +90.4$\pm$4.2\%. Within the uncertainties, the degree of polarization between these two epochs is similar.}

%\newline DAF. I think we capture this behavior in the first paragraph of section 2.2.1 by summarizing the original detections from our first paper. I see the argument, however, for wanting to report the pol properties with the same analysis we do for the light curves and spectra. Opinions? SDH: The 16\% difference between 94 and 78\% on the two epochs is just a tad greater than the combined uncertainties in quadrature, so that in itself is not noteworthy. But perhaps the superbright 5 min scan's 90.4 pm 4.2\% is worth noting compared to the 70.6 pm 10.8\% for July 10 without that scan. 

%\textcolor{red}{Note we don't really say anything about the handiness and the degree of polarization for these three data cuts. Any thoughts on how to do this? We do mention V/I in the first paragraph. Maybe that is enough. Related to that, I think the reader would rather that we use error bars rather than S/N for the V/I values.}

The second dataset was collected during the 2021 discovery outburst of the black hole candidate MAXI\,J1803$-$298 \citep{2022ApJ...926L..10M,2022MNRAS.516.2074F,Espinasse2025}. MeerKAT undertook a weekly monitoring campaign at 900--1670 MHz from 2021 May 04, 2021 to Deccember 18 for a total of 30 epochs, each about $\sim$0.2-0.3 hrs in duration \citep{2021ATel14607....1E}. All observations were obtained in full polarization mode, 8 seconds integration time per visibility (dump time) and with a total (un-flagged) bandwidth of 856 MHz split evenly into 32768 frequency channels. The scans were typically alternated between the target and phase calibrator (J1833$-$2103 or J1830$-$3602). J1939$-$6342 was used as primary calibrator for flux and bandpass calibration. 
\textsc{OxKAT}\footnote{\url{https://github.com/IanHeywood/oxkat}}, a semi-automated routine, was used for the data calibration and imaging. In particular, the Common Astronomy Software Application\footnote{\url{https://casa.nrao.edu/}} \citep[CASA;][]{2022casa} is called for phase correction, antenna delays and bandpass correction while \textsc{Tricolour} \footnote{\url{https://github.com/ratt-ru/tricolour}} \citep{2022hugo} is called for interference flagging. \textsc{OxKAT} calls \textsc{WSClean} \citep{wsclean} for imaging and \textsc{Cubical}\footnote{\url{https://github.com/ratt-ru/CubiCal}} for self-calibration \citep{cubical}. 
 Before imaging, we averaged together every 32 channels in order to decrease the size of each dataset. Imaging of the data was carried out using a Briggs weighting scheme (robust=$-$0.3). 
 %The average rms noise of a single epoch is 20 $\mu$Jy/beam and the median synthesized beam is 6.2". 

%The observation method and the data analysis method that we followed was identical to that described in \cite{2024MNRAS.527.9359H}. 

%\textcolor{red}{Lilia, feel free to modify or expand on the above text. This next paragraph is where Scott and Lilia will write up what they find. Do we confirm the basic behavior above and do we learn anything new? I suggest that we do not put all 30 epochs in Table 1 but maybe we add it like the 41 ASKAP epochs, and reference a light curve, and/or just adding the 3sigma detections?}

\begin{figure}[!ht]
\centering
\includegraphics[width=0.98\textwidth,angle=0,trim=0cm 0cm 0cm 0cm]{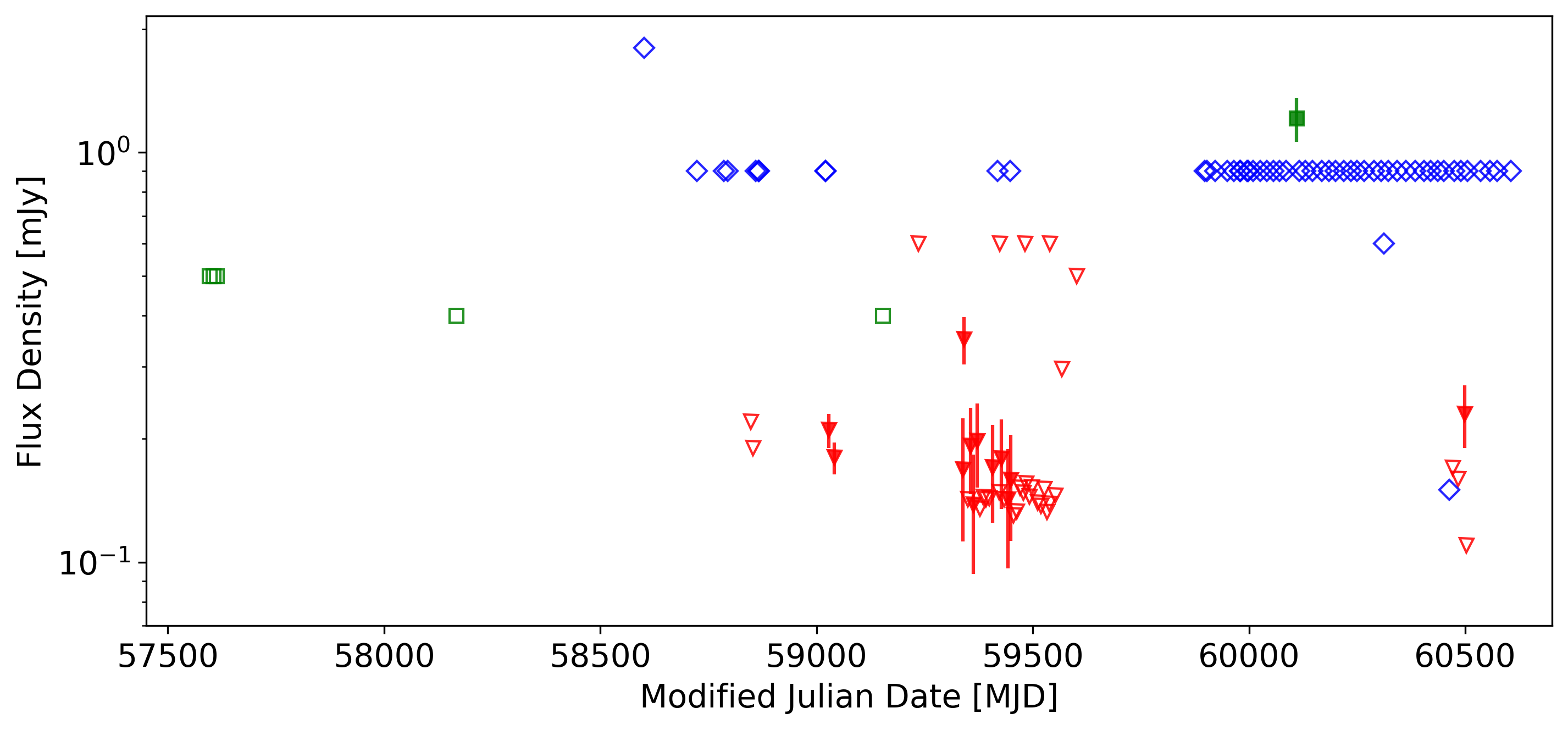}
\caption{An 8-year radio light curve of \hd from the data in Table \ref{tab:radio}. Triangles (red) are data taken at frequencies between 1 and 2 GHz. Diamonds (blue) are for frequencies below 1 GHz, while squares (green) are for frequencies above 2 GHz.
Measured flux densities are indicated with filled symbols with 1-$\sigma$ error bars, while 3-$\sigma$ limits are open symbols.}
\label{fig:master_lightcurve}
\vspace{1cm}
\end{figure}

The measured flux density and rms values for each epoch are listed in Table \ref{tab:radio} and plotted in Fig.\,\ref{fig:master_lightcurve} as a function of the Modified Julian Date (MJD). In this and all other observations, where \hd was not at the phase center, we have applied primary beam corrections to these values. Of the 30 epoch 1.28-GHz images, radio emission is detected coincident with \hd in about half the epochs, but is brightest during the 2021 May 07 epoch. The flux densities and rms noise measured at the position of the radio source for each epoch are listed in Table \ref{tab:radio}.

%half are detected at least marginally upon visual inspection. The 2021 May 07 epoch is the brightest with a flux density at the peak pixel of 0.377 $\pm$ 0.047 mJy ba$^{-1}$ \sout{(8.0$\sigma$)}, after applying a correction factor of 2.33 for primary beam attenuation. {\bf The local noise and resolution are 0.05 mJy ba$^{-1}$ and 6.7 $\times$ 5.3"} A Gaussian fit yields a nominally resolved source with peak position 18$^h$ 05$^m$ 26.49$^s$ $\pm$ 0.02$^s$, $-$29$^\circ$ 29${^\prime}$ 53.5$^{\prime\prime}$ $\pm{0.4}^{\prime\prime}$. \sout{The peak and integrated flux densities are 0.35 $\pm$ 0.04 mJy ba$^{-1}$ and 0.48 $\pm$ 0.09 mJy, respectively.} However, imaging together seven epochs with greater than 3$\sigma$ detections yields a clearly unresolved 0.23 $\pm$ 0.02 mJy (10$\sigma$) source. \sout{The local noise and resolution of the combined epochs (May 7) whole band images are 0.02 (0.05) mJy ba$^{-1}$ and 6.3 $\times$ 5.5" (6.7 $\times$ 5.3"), respectively.} The flux densities {\bf or limits} for each epoch are listed in Table \ref{tab:radio}. Since only a few of the detections are over 4$\sigma$, the flux densities of each were measured at the May 7 peak where the source position is best determined.
% Lilia's epoch images have circular beams. I remade May 7 without enforcing that and got somewhat better fit. (Less resolved and less uncertainty.) 

While \hd was never as radio bright in 2021 as it was in 2020, we attempted to constrain the spectral index. Imaging the seven combined epochs when the source was detected in two halves centered at 1.07 and 1.50 GHz, the spectral index is $-$0.54$\pm$0.48. Likewise, if  we break the data from the bright May 07 epoch into a low and high channels the spectral index is +0.02$\pm$0.70. Thus, although the quality of the in-band fits is poor, it appears therefore that the 2021 radio spectrum is not as steep as the 2020 observations. We have no significant constraints on circular or linear polarization for these 2021 observations because a calibrator source was not included for this purpose.

Summarizing, in 2020 we detected L band ({hereafter 1.5 GHz}) emission that is highly circularly polarized (V/I=+90\%) with a sharp break above 1 GHz, after which the spectrum turns unusually steep ($\alpha=-5$). The radio source is significantly variable month to month, during which the source is active on timescales of several hours or longer. While the source is active there are significant hour-to-hour fluctuations in flux density, and during the brightest ``flares" we see evidence for fluctuations on 30-s to 15-minute timescales. The high fractional polarization, its polarity, and the steep spectrum appear to be constant while the source is active. This radio activity appears to continue on week-long timescales in 2021 based on a series of shorter observations, although the flux densities are not as bright, and the spectrum does not appear to be as steep.

%\begin{figure}[!ht]
%\centering
%\includegraphics[width=0.98\textwidth,angle=0,trim=0cm 0cm 0cm 0cm]{masterspectrum.png}
%\caption{Stokes RR spectrum with broken power law model fit of brightest 5-min 2020 July 10 scan (upper red) and average Stokes RR spectrum (lower red) from combination of all 1-hr pointings. Black points indicate Stokes I detections from GMRT on 2024 July 7, ThunderKAT on 2021 May 7, and VLASS on 2023 June 14. Open triangles indicate Stokes I upper limits from the telescopes listed in Table \ref{tab:radio}.}
%\label{fig:master_spectra}
%\vspace{1cm}
%\end{figure}

\startlongtable
\begin{deluxetable}{lllrrr}
\tablecaption{Time-Ordered Summary of Recent and Archival Radio {Measurements}\label{tab:radio}}
\tablehead{\colhead{Obs. Date} & MJD & \colhead{Telescope} & \colhead{Frequency} & \colhead{Flux Density} & Reference\\
\colhead{(YY-MM-DD)} & \colhead{}  & \colhead{} & \colhead{(MHz)} & \colhead{({mJy)}} & \omit}
\startdata
{1996-05-10} & 50213 & VLA & 1400 & $<$2.2 & NVSS \\ %AC308 3-sigma on skyview image is 1.5 mJy. All other limits changed to 3 sigma.
2010-04-28 & 55314 & GMRT & 150 & $<$14 & TGSS\\ % based on skyview image. Archive says obs on 28 APR 2010.
2016-07-27 & 57596 & VLA & 3000 & $<0.5$ & VLASS \\ % pilot QL image
2016-08-05 & 57605 & VLA & 3000 & $<0.5$ & VLASS \\ % pilot QL image
2016-08-13 & 57613 & VLA & 3000 & $<0.5$ & VLASS \\ % pilot QL image
2018-02-18 & 58167 & VLA & 3000 & $<$0.4 & VLASS \\ % QL image
2018-05-22 & 58260 & VLA & 340 & $<$6.6 & VLITE\\
2019-04-28 & 58601 & ASKAP & 888 & $<$1.8 & RACS-low \\ % badwidth 288 MHz
2019-11-08 & 58795 & ASKAP & 888 & $<$0.9 & RACS-low \\
2019-12-31 & 58848 & MeerKAT & 1284 & $<$0.22 & This Paper\\ % pointing is 0.9 deg from source. consider not including at all).
2020-01-04 & 58852 & MeerKAT & 1284 & $<$0.19 & This Paper\\ % same note as above
2020-06-20 & 59020 & ASKAP & 888 & $<$0.9 & VAST-low \\
2020-06-28 & 59028.88 & MeerKAT & 1284 & $0.21 \pm 0.02 $ & This Paper \\ % not enough signal in high half of band. But extrap from low half yields ~0.2. ALSO, EXTRAP FROM 1144 MHZ POINT WITH PRESUMED -5.5 ALPHA YIELDS 0.2 TOO. Same result using PBCOR on whole band (no reamp). Imaging just 1st and 2nd IFs in RR shows 6/28 is 25% higher than July 10 nearest pointing.
2020-07-10 & 59040.88 & MeerKAT & 1284 & $0.18 \pm 0.02$ & This Paper\\ % nearest pointing. Whole band image. Consistent with a broken spectrum fit too. There are two other July 10 pointings at 0.5 and 0.8 deg, all included in light curve.
2020-10-31 & 59153 & VLA & 3000 & $<$0.4 & VLASS \\ % QL image
2021-01-22 & 59236 & ASKAP & 1368 & $<$0.6 & RACS-mid \\ % bandwidth 144 MHz
2021-05-04 & 59338.90 & MeerKAT & 1284 & 0.19$\pm$0.06 & This Paper\\
2021-05-07 & 59341.13 & MeerKAT & 1284 & 0.38$\pm$0.05 & This Paper\\
2021-05-15 & 59349.93 & MeerKAT & 1284 & 0.12$\pm$0.05 & This Paper\\
2021-05-22 & 59356.05 & MeerKAT & 1284 & 0.19$\pm$0.05 & This Paper\\
2021-05-27 & 59361.98 & MeerKAT & 1284 & 0.12$\pm$0.04 & This Paper\\
2021-06-05 & 59370.92 & MeerKAT & 1284 & 0.22$\pm$0.05 & This Paper\\
2021-06-12 & 59377.98 & MeerKAT & 1284 & 0.08$\pm$0.05 & This Paper\\
2021-06-19 & 59384.94 & MeerKAT & 1284 & 0.02$\pm$0.05 & This Paper\\
2021-06-27 & 59392.06 & MeerKAT & 1284 & 0.03$\pm$0.05 & This Paper\\
2021-07-04 & 59399.93 & MeerKAT & 1284 & 0.14$\pm$0.05 & This Paper\\
2021-07-12 & 59407.05 & MeerKAT & 1284 & 0.13$\pm$0.05 & This Paper\\
2021-07-26 & 59421.72 & MeerKAT & 1284 & 0.02$\pm$0.05 & This Paper\\
2021-07-29 & 59424 & ASKAP & 1368 & $<$0.60 & VAST-mid \\
2021-07-31 & 59426.86 & MeerKAT & 1284 & 0.15$\pm$0.04 & This Paper\\
2021-08-07 & 59433.84 & MeerKAT & 1284 & 0.10$\pm$0.05 & This Paper\\
2021-08-15 & 59441.89 & MeerKAT & 1284 & 0.12$\pm$0.05 & This Paper\\
2021-08-22 & 59448.65 & MeerKAT & 1284 & 0.14$\pm$0.05 & This Paper\\
2021-08-28 & 59454.65 & MeerKAT & 1284 & 0.05$\pm$0.04 & This Paper\\
2021-09-05 & 59462.72 & MeerKAT & 1284 & 0.01$\pm$0.04 & This Paper\\
2021-09-13 & 59470.62 & MeerKAT & 1284 & 0.05$\pm$0.05 & This Paper\\
2021-09-20 & 59477.65 & MeerKAT & 1284 & $-$0.02$\pm$0.05 & This Paper\\
2021-09-25 & 59482 & ASKAP & 1368 & $<$0.60 & VAST-mid \\
2021-09-27 & 59484.57 & MeerKAT & 1284 & 0.09$\pm$0.05 & This Paper\\
2021-10-04 & 59491.63 & MeerKAT & 1284 & $-$0.07$\pm$0.05 & This Paper\\
2021-10-09 & 59496.63 & MeerKAT & 1284 & $-$0.08$\pm$0.05 & This Paper\\
2021-10-23 & 59510.53 & MeerKAT & 1284 & $-$0.02$\pm$0.05 & This Paper\\
2021-10-31 & 59518.61 & MeerKAT & 1284 & 0.08$\pm$0.05 & This Paper\\
2021-11-08 & 59526.65 & MeerKAT & 1284 & 0.08$\pm$0.05 & This Paper\\
2021-11-14 & 59532.50 & MeerKAT & 1284 & 0.13$\pm$0.04 & This Paper\\
2021-11-19 & 59537.69 & MeerKAT & 1284 & $-$0.01$\pm$0.05 & This Paper\\
2021-11-21 & 59539 & ASKAP & 1368 & $<$0.60 & VAST-mid \\
2021-12-04 & 59552.44 & MeerKAT & 1284 & 0.15$\pm$0.05 & This Paper\\
2021-12-18 & 59566.46 & MeerKAT & 1284 & $<$0.30 & This Paper\\
2022-01-23 & 59602 & ASKAP & 1655 & $<$0.5 & RACS-high \\ % bandwidth 200 MHz
2023-06-14 & 60109.31 & VLA & 3000 & $1.57 \pm 0.16$ & VLASS \\ % QL image is 1.21 pm 0.15.
2023-06-14 & 60109 & VLA & 340 & $<$60 & This Paper \\
2024-01-03 & 60312 & ASKAP & 943 & $<$0.6 & RACS-low\\ % bandwidth 288 MHz
2024-06-02 & 60463 & GMRT & 650 & $<$0.15 & This Paper\\ % 1.0x flux scale correction
2024-06-09 & 60470.98 & GMRT & 1260 & 0.04$\pm$0.06 & This Paper\\ % 1.64x flux scale correction
2024-06-21 & 60482.91 & GMRT & 1260 & -0.01$\pm$0.06 & This Paper\\ % 1.34x correction
2024-07-07 & 60498.66 & GMRT & 1260 & $0.23 \pm 0.04$ & This Paper\\ % 1.20x correction
2024-07-11 & 60502.75 & GMRT & 1260 & $0.14\pm0.05$ & This Paper\\ % 1.2x corr. marginal 3 sigma detec. stronger in 1st half?
\hline
2016-2020; 5 epochs & \omit & VLA       & 3000  & $<$0.20 & VLASS\\ % average of two nondetection images and 3 pilot ones
2019-2024; 50 epochs & \omit & ASKAP     & 887  & $<$0.20 & VAST-low\\  % 50 12-min scans thru 2024-10-22 
2021; 14 epochs & \omit & MeerKAT & 1284 & $<$0.043 & This Paper\\ %14 12-min scans. (Two in June, rest Aug-Nov)
2024; 3 epochs & \omit & GMRT     & 1260 & $<$0.085 & This Paper\\ % June 10, 22, July 11 image from combined uv data. 1.31x flux scale correction
\enddata
\tablecomments{(a) The last four rows combine data from multiple epochs in a deeper search to constrain the quiescent flux density. (b) The 1284-MHz fluxes for the 2021 MeerKAT epochs were measured at the 2021 May 7 peak position. (c) The 1260-MHz fluxes for the 2024 GMRT epochs were measured at the 2024 July 7 peak position.}
\end{deluxetable}

\subsubsection{Giant Metrewave Radio Telescope}\label{sec:gmrt}

We observed \hd with the upgraded Giant Metrewave Radio Telescope (uGMRT) at 550--900 MHz (Band 4) and at 950--1460 MHz (Band 5) for 4 epochs in June and July of 2024.
%on 2024 June 2 and at }  on 2024 June 10, June 22, July 7 and July 11 for 1 hour on each day, including 30 min on-source and calibration overheads (project DDT C352). 
All data were collected in spectral-line mode using the wide-band back-end with a bandwidth of 400 MHz, 8192 frequency channels, and an integration time of 5.3 seconds. We observed the source 3C\,286 as bandpass and absolute flux density calibrator during all observations except June 10, when we used instead the calibrator source 3C\,48. The sources 1830$-$360 and 1751$-$253 were observed as phase calibrators. We reduced the data using the Source Peeling and Atmospheric Modeling \citep[\texttt{SPAM};][]{2009A&A...501.1185I} pipeline\footnote{http://www.intema.nl/doku.php?id=huibintemaspampipeline}. 
For each observation, the full-band dataset was first divided into 6 narrower channels. We then calibrated each channel independently with the \texttt{SPAM} pipeline, adopting a standard calibration scheme that consists of bandpass and complex gain calibration, as well as flagging of radio frequency interference. The flux density scale was set using
\cite{2012MNRAS.423L..30S}. After initial calibration, we applied phase self calibration to the target data using our MeerKAT image as an initial sky model. For each observation, we imaged the final self-calibrated channels together using joint-channel deconvolution in \texttt{WSClean} \citep{wsclean} and produced final images at the central frequencies of 700 MHz (Band 4) and 1260 MHz (Band 5) using a Briggs weighting of 0 and an inner uv-plane cut at 10\,k$\lambda$. 

%The final image at 700 MHz has a resolution of $3^{\prime\prime}.1\times7^{\prime\prime}.5$ and a noise of $\sim 50$ $\mu$Jy ba$^{-1}$. The Band-5 images have {\bf typical}resolutions and local noise levels of {\bf $\sim 2^{\prime\prime}\times5^{\prime\prime}$ and $\sim 40$ $\mu$Jy ba$^{-1}$.} \sout{$1^{\prime\prime}.4\times7^{\prime\prime}.5$ and $\sim 40$ $\mu$Jy ba$^{-1}$ (June 10), $1^{\prime\prime}.6\times6^{\prime\prime}.3$ and $\sim 39$ $\mu$Jy ba$^{-1}$ (June 22), $2^{\prime\prime}.0\times3^{\prime\prime}.6$ and $\sim 34$ $\mu$Jy ba$^{-1}$ (July 7), $1^{\prime\prime}.7\times3^{\prime\prime}.7$ and $\sim 34$ $\mu$Jy ba$^{-1}$ (July 11), respectively.} 
%{\tt Scott, please cross check the final noise levels - these are the ones given by wsclean for the whole imaged field, but you may want to replace them with the local rms. SDH: I updated them including 5-20\% flux scale corrections.}

An unresolved source is detected at 1260 MHz coincident with \hd
%at RA 18$^h$ 05$^m$ 26.44$^s$, DEC $-$29$^\circ$ 29$^{\prime}$ 53.7$^{\prime\prime}$ 
in the July 7 observation. The signal-to-noise ratio was insufficient to derive any meaningful constraints on an in-band spectrum.
To search for possible variability, we imaged six 5-minute intervals, with the fourth showing a significant increase to $0.40 \pm 0.09$ mJy. Imaging each half of the observation results in a $2\times$ increase from $0.17 \pm 0.05$ to $0.34 \pm 0.06$ mJy.
%field sources used: 18 05 52.68, -29 28 10.35; 18 05 59.32, -29 31 02.2; 18 05 16.01,-29 25 54.0. ALL YIELD CONSISTENT 1.2x FLUX SCALE CORRECTION BASED ON WHOLE BAND GMRT TO WHOLE BAND MEERKAT COMPARISON; CENTRAL FREQS VERY SIMILAR: 1.26 AND 1.28 GHZ.
%Note that a really bright resolved source about 8.5 south is anonomously super steep using the half bands---not in agreement with meerkat at all. Something must be wrong with pbcor in high half band there, although should be well within half power still. Another source about same distance east totally gone in high half, but shouldn't be. OH WAIT: MEERKAT BEAM IS 8" whereas GMRT high half beam is 3.5 x 1.75" so maybe much of the flux is resolved out in the gmrt high half image. But the low half beam is similar 4 x 2. IGNORE?
None of the other 2024 June and July GMRT observations convincingly detect the source.
%although there is a possible 3$\sigma$ faint detection on July 11 {\bf at 1260 MHz}

The GMRT detection and upper limits are listed in Table \ref{tab:radio} and plotted in Fig.\,\ref{fig:master_lightcurve}. While these data do not give useful constraints on either the spectral index or circular polarization properties of the radio emission, they provide an important verification of the MeerKAT results in terms of the ON/OFF duty cycle, the average flux density, and the appearance of significant variability on timescales of minutes.

%Not even sure combining the nondetections would be a true quiescent  flux if low level activity is lasting weeks.  In any case, ignoring 3 sigma bumps in 6/22 and 7/11, combining them with 6/10 nondetection leads to 0.085 mJy 3 sigma upper limit (after 1.31x flux scale correction). 

\subsubsection{Archival Radio Data}\label{sec:archive}

In an effort to better constrain the spectrum and light curve behavior of \hdp, we searched previously published radio surveys and existing on-line archives. Radio surveys included the TGSS \citep[150 MHz,][]{2017A&A...598A..78I}{}, GLEAM \citep[200 MHz,][]{2017MNRAS.464.1146H}, VLITE \citep[340 MHz,][]{2016SPIE.9906E..5BC}{}{}, SUMSS \citep[843 MHz,][]{1999AJ....117.1578B}, RACS-low \citep[888 MHz,][]{2021PASA...38...58H}{}, RACS-mid \citep[1368 MHz,][]{2023PASA...40...34D}{}{}, and RACS-high \citep[1655 MHz,][]{racshigh}, NVSS \citep[1.4 GHz,][]{1998AJ....115.1693C}{}{}, and the six VLASS epochs \citep[3 GHz,][]{2020PASP..132c5001L}{}{}. 

%The observing date, telescope, frequency and flux density (or 3$\sigma$ limits) from each of these archival datasets are listed in Table \ref{tab:radio}. 
%When raw data was available, we downloaded and re-reduced the archival data directly, otherwise...

For the VLASS and the MeerKAT datasets we calibrated the raw visibilities taken from the archive, while for the remainder, we downloaded and analyzed images produced by the survey team. For example, from the ASKAP Variables and Slow Transients survey \citep[VAST;][]{2021PASA...38...54M} we found fifty 12-minute archival ASKAP observations centered at 888 MHz, significantly overlapping the low-end of the MeerKAT wide-band. The majority of the archival values in Table \ref{tab:radio} are non-detections and for these we provide 3-$\sigma$ upper limits.

%None detect the source, and all have an rms noise level of $\sim$0.3 mJy ba$^{-1}$. We include the 20 June 2020 observation, 8 days prior to the June 2020 MeerKAT detection, as a representative upper limit in Table \ref{tab:radio}. After averaging the 50 images, the resulting Stokes I non-detection has a 3$\sigma$ upper limit $\sim$50\% lower than the $\sim0.5$ mJy 900 MHz end of the mean MeerKAT spectrum (Fig. \ref{fig:master_spectra}). The majority of the archival values in this table are non-detections and these values set limits on the quiescent flux density of \hdp. We plot the best limits at each frequency on Fig. \ref{fig:master_spectra}. At 3 GHz we added in quadrature 2018 and 2020 values from Table \ref{tab:radio} for an upper limit of 0.27 mJy.
%SDH: The sentence above comparing the 50-epoch average to the low end of the meerkat spectrum is awkward since the 50-avg is Stokes I, but the figure plots RR, which levels off at 1 mJy. (Stokes I then is about 0.5 mJy, and the 50-avg is 0.2 mJy.) 

There is one detection among these archival data. A 2-4 GHz (S-band) image made as part of the third epoch of the Very Large Array Sky Survey (VLASS) on June 14, 2023 whereas it was undetected in any prior observations (see Tables \ref{tab:radioobs} and \ref{tab:radio}). As VLASS is carried out in scan mode (3.31 arcmin/sec), we used the VLA CASA-pipeline to reduce and make mosaic images of the June 14, 2023 data choosing two sets of seven pointings, $\sim$1 minute apart, nearest to \hdp. The flux density and rms in Table \ref{tab:radio} is based on an image made from the combination of the two sets. The next set of pointings 7 minutes later 
%at declination $-29^\circ$ 38$^\prime$ and 
yields no detection with a 3-$\sigma$ limit of 1.1 mJy, perhaps indicating a reduction in activity. The in-band spectrum is flat across the 2-4 GHz frequency range with some evidence for a slight drop at the high and low frequency edges. A power-law fit to the spectral index gives $\alpha=+0.85\pm{0.86}$, inconsistent with the extremely steep spectrum seen by MeerKAT (\S\ref{sec:meerkat} and Fig.\,\ref{fig:meerkat}).

Summarizing, these archival data show emission centered at a frequency of 3 GHz whose spectrum, duration and polarization are materially different than the 1.5-GHz emission (\S\ref{sec:meerkat}). While the rest of the archival data are non-detections, the absence of ASKAP detections at 888 MHz (plus TGSS and VLITE non-detections) provides some support that the steep $\alpha=-5.5$ MeerKAT spectrum turns over below 1 GHz (see Fig.\,\ref{fig:meerkat}). Alternatively, as we lack simultaneous broadband radio observations, the source may not have been radio variable during any of the 50 ASKAP epochs at 888 MHz (Fig.\,\ref{fig:master_lightcurve}). The best limits on quiescent emission in or near 1.5 GHz come from averaging the 14 ThunderKAT 2021 nondetections, the three 2024 GMRT nondetections, and the 50 ASKAP 2019-2024 nondetections. These $3\sigma$ limits are listed at the bottom of Table \ref{tab:radio}.

% Simona actually imaged the 3 nondetection GMRT epochs uvdata together as opposed to averaging 3 images. The 0.085 mJy is from that.  The ASKAP average is obtained by averaging the 41 images.

\subsubsection{Radio Periodicity}\label{sec:period}

\begin{figure}
\centering
\includegraphics[width=0.85\textwidth,angle=0,trim=0cm 0cm 0cm 0cm]{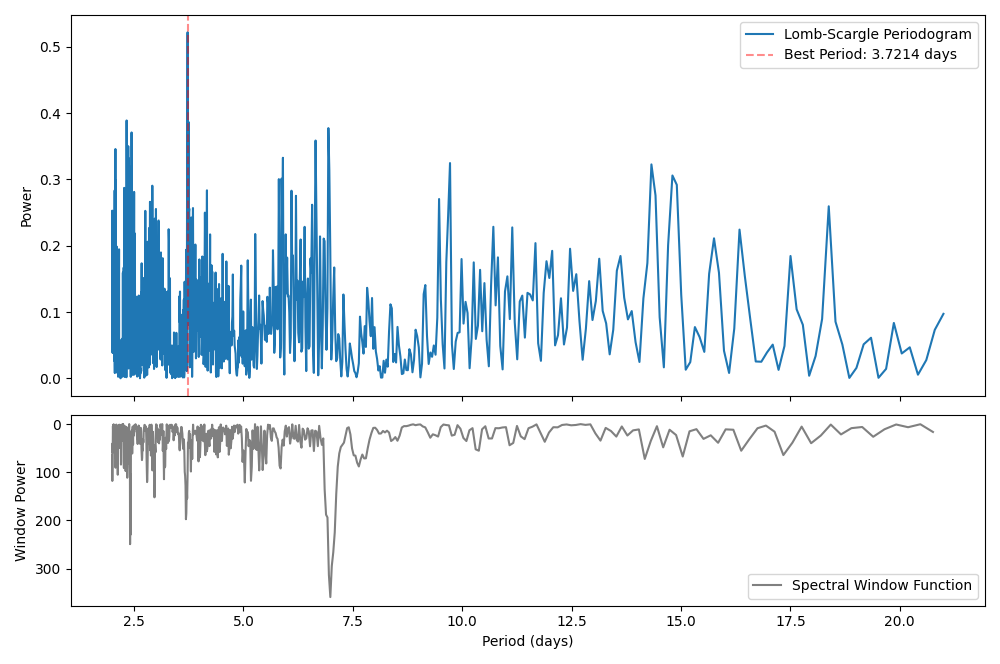}
\caption{A Lomb-Scargle periodogram of the 1.5-GHz radio measurements taken toward \hd from 2019 to 2024. A peak with a period of 3.7 days is detected with a signal-to-noise of 7.3. The corresponding spectral window function for these data is also shown.}
\label{fig:radioperiod}
\vspace{1cm}
\end{figure}

The emission activity at 1.5 GHz appears to be periodic, not stochastic. We calculated a Lomb-Scargle periodogram from astropy \citep{2022ApJ...935..167A}, using all epochs in Table \ref{tab:radio} with MJD, flux and rms values. The resulting periodogram in Fig.\,\ref{fig:radioperiod} shows a peak at P=3.72$\pm$0.05 days (S/N=7.3). Assuming there is no periodic signal in these data, the false alarm probability of such a peak appearing at random in the period range between 1 and 20 days is 6$\times{10}^{-4}$ \citep{2008MNRAS.385.1279B}.
%Applying bootstrapping to these data indicates that the period uncertainty is rather large ($\pm$2.2 days). 

As this period is close enough to a multiple of one week, we subjected these data to a number of tests. First
we looked at the histogram of time differences between observations. There are two modest peaks near 7 and 14 days, likely reflecting the near-weekly MeerKAT monitoring between May and December 2021, but the overall distribution is consistent with these data being sampled at uneven intervals. Next we estimated the spectral window function by calculating the discrete Fourier transform of the sample distribution. Between 2 and 21 days the largest peaks in order are at 7.0, 2.4 and 3.6 days; suggesting that we are more sensitive to periods or harmonics close to these values (Fig.\,\ref{fig:radioperiod}). Finally we searched for the same period using a light curve from a nearby (non-variable) continuum point source. We find no evidence of any periodicity at the same period seen for \hdp.  

\begin{figure}
\centering
\includegraphics[width=0.85\textwidth,angle=0,trim=0cm 0cm 0cm 0cm]{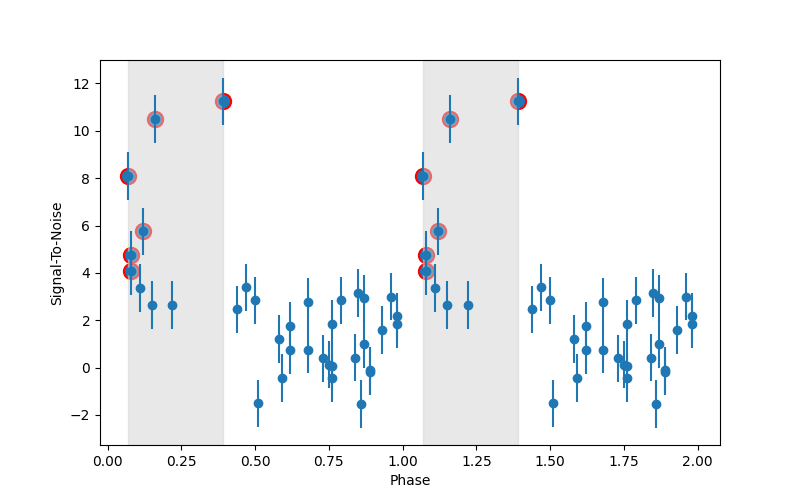}
\caption{A phase-folded radio light curve at 1.5 GHz toward \hd derived, using the best-fit Lomb-Scargle period P=3.72 days. The (light grey) shaded vertical regions are near show the clustering of our most significant detections (S/N$\geq$4). These six points are indicated with larger (red) circles than our marginal detections and limits.}
\label{fig:phasefold}
\vspace{1cm}
\end{figure}

The resulting phase-folded light curve for \hd is also shown in Fig.\,\ref{fig:phasefold}. These data can be fit with a simple sinusoidal ($\chi^2_r=$1.9) but an ON-OFF boxcar function is also possible. We note that our most significant detections (S/N$\geq$4) are clustered in a phase interval 0.30.
%, consistent with the f=25\% duty cycle derived independently in \S\ref{sec:archive}. 
This can be a useful quantity to constrain the origin of the emission \citep{2023MNRAS.524.6267K}, and suggests that the 1.5-GHz emission may be beamed. The heterogeneous sampling precludes strong claims about either periodicity or beaming. Future observations will be needed to better constrain both the period and the shape of the light curve before stronger claims can be made. 

An alternate estimate of the duty cycle (f) without regard to periodicity is simply the fraction of the time that \hd is active enough to be detected.  At 3 GHz, we derive f~$\simeq{17}\%$, based on one detection out of 6 VLASS similar epochs. At 1.5 GHz this is complicated by the diverse set of observations. The duration of each epoch ranges from 12 minutes to several hours. Likewise, the sensitivity varies by almost an order of magnitude and some of the observing programs have a weekly or near monthly cadence, while others consist of several randomly scheduled epochs. Even the center frequency ranges from 0.89 GHz to 1.4 GHz. For example, the 2019 and 2020 MeerKAT observations consisted of twelve 4.5-minute scans in each of several different pointings, on four days spread over 9 hr observing periods, while the 30 epochs in 2021 each consisted of 12-minute scans. The fifty VAST-low and the various RACS-low and RASC-high epochs would have detected only the brightest MeerKAT fluctuations. The GMRT and the 2020 and 2021 MeerKAT observing programs provide the best constraints on the duty cycle. Just counting detections versus non-detections for these three individual observing runs, the value of f varies between 19\% to 30\%. If we account for the time-on-sky for all three programs we derive f=27\%. At 1.5 GHz we will adopt f=25\% but recognize that it is not well constrained.

%but an unfortunate sampling gap in the ON interval precludes any further analysis. 

\subsection{X-Ray Observations}\label{sec:xrays}

\hd 
%has been detected at X-ray wavelengths on several occasions. It 
is in the {\em Chandra} Source X-ray Catalog \citep[CSC;][]{2010ApJS..189...37E} as 2XCO\,J180526.4$-$292952, and also
%it is found 
in the XMM-Newton Serendipitous Source Catalog \citep[4XMM-DR13;][]{2020A&A...641A.136W} as XMM\,J180526.3$-$292951. We list the catalog fluxes (f$_x$) and S/N values in Table \ref{tab:xray} along with the dates, observing identification, bands and integration times. In the XMM light curves there is no indication of any significant short-term variations in the flux during either of observing session.

%\hd has been detected at X-ray wavelengths on several occasions. It is in the {\em Chandra} Source X-ray Catalog \citep[CSC;][]{2010ApJS..189...37E} as 2XCO\,J180526.4$-$292952, visible on a 1.2 ksec integration on 2007-10-23 (OBSID=7541) with 3.3$\sigma$ significance. The CSC v2.1 lists the aperture-corrected flux f$_x$ in the 0.1-10 keV band as 1.79$\times{10^{-13}}$ erg s$^{-1}$ cm$^{-2}$. In the XMM-Newton Serendipitous Source Catalog \citep[4XMM-DR13;][]{2020A&A...641A.136W} \hd is coincident with XMM\,J180526.3$-$292951, observed on 2016-09-06 (ObsID=0784100101) and 2016-09-27 (ObsID=0782770101) for $\sim$24 ks and $\sim$50 ks, respectively. A source is detected with high significance with flux in the 0.2-12 keV band of 2.71($\pm{0.28})\times{10^{-13}}$ erg s$^{-1}$ cm$^{-2}$ and 1.59($\pm{0.15})\times{10^{-13}}$ erg s$^{-1}$ cm$^{-2}$, respectively, or a mean flux of 1.84($\pm{0.13})\times{10^{-13}}$ erg s$^{-1}$ cm$^{-2}$. There is no indication of any significant short-term variations in the flux during either of 2016-09-06 and 2016-09-27 observing sessions.

\begin{table}[!ht]
    \caption{Summary of Archival X-ray Observations of \hdp}\label{tab:xray}
    \centering
    \begin{tabular}{cccccccr}
        \hline
        Telescope & Obs. ID & Date & Band & Integration & f$_x$ & L$_x$ & S/N \\
        \omit & \omit & (YYYY-MM-DD) & (keV) & (ks) & (erg s$^{-1}$ cm$^{-2}$) & (erg s$^{-1}$) & \omit\\
        \hline
        \hline
        Chandra & 7541 & 2007-10-23 & 0.1-10 & 1.2 & 1.79$\times{10}^{-13}$ & 2.45$\times{10}^{28}$ & 3.3$\sigma$ \\
        XMM & 0784100101 & 2016-09-06 & 0.2-12 & 24 & 2.71$\times{10}^{-13}$  & 3.71$\times{10}^{28}$ & 9.7$\sigma$ \\
        XMM & 0782770101 & 2016-09-27 & 0.2-12 & 50 & 1.59$\times{10}^{-13}$ & 2.19$\times{10}^{28}$ & 10.7$\sigma$ \\
        \hline
    \end{tabular}
    \vspace{1cm}
\end{table}

Despite slightly different energy bands, the f$_x$ measured by \emph{Chandra} and the mean f$_x$ from \emph{XMM-Newton} are comparable, suggesting that the X-ray emission from \hd is reasonably constant on decade-long timescales. Using f$_x$ and the distance (d) we derive the X-ray luminosity (L$_x=4\pi$d$^2$ f$_x$).
%of 2.52$\times{10}^{28}$ erg s$^{-1}$.

%\textcolor{red}{Notes:  Talk to some experts. What are K and M stars X-ray fluxes normally? can we fit the spectra with something like XPEC (model apec), i.e. to get the plasma temperature for a single photoelectrically absorbed optically thin plasma model (i.e. like Callingham et al 2021). We have enough to measure Lx now but later we may want to get a handle on the T$_{plasma}$ to rule out models.}
 
% Notes: For Chandra I used CSCView to get the basic properties and the Source Observation Summary. It does give us RA/Dec that we might be abl eto use. 18 05 26.43,-29 29 52.30,     with errors of +/-1.01 arcsec.
% For XMM the source is at: https://xcatdb.unistra.fr/4xmmdr13/xcatindex.html?srcid=207841001010002
% There are a lot of details there, including light curves, spectra and positions. 

\subsection{Optical/Near Infrared Observations}\label{sec:optical}

\hd is the bright primary of a nearby high proper motion visual binary system separated by 4.8$^{\prime\prime}$ \citep{2021AA...649A...1G, 2021AJ....161..147B}. This system does not appear to be a member of any nearby young association \citep{2018ApJ...856...23G}. Despite both \hd and HD\,317101B being rather bright (G$\simeq$10 mag), neither have received much attention in the literature prior to the {\it Gaia} mission. Since the radio emission clearly originates from the direction of \hdp, we focus our discussion on this star.

We list the basic optical/infrared parameters that are known for \hd in Table \ref{table:optical}. For this parallax value, the distance modulus to convert observed magnitudes to absolute is $\mu=-2.631 \pm 0.009$. The stellar parameters listed in Table \ref{table:optical} (e.g., mass, T$_{\rm eff}$, age, log\,$g$) are derived from the {\it Gaia} stellar catalog under the source ID as DR3\,4050315620888089472 \citep{2023AA...669A.104K} and the TESS Input Catalog (TIC) as TIC\,407888149 \citep{2019AJ....158..138S}. If \hd was a single star we estimate that it would be a K6.5V star based on its {\it Gaia} data release 3 (DR3) absolute magnitude (M$_G$) and colors (BP-RP), and using the spectral notes\footnote{\url{https://github.com/emamajek/SpectralType/blob/master/EEM_dwarf_UBVIJHK_colors_Teff.txt}} from \citet{2013ApJS..208....9P}.

There is good evidence, however, to suggest that \hd has an unseen companion star. The first indication comes from the {\it Gaia} re-normalized unit weight error (RUWE), a reduced $\chi^2$-like statistic that is a quality measure of the single-star astrometric solution. The RUWE=19.7 for \hd is an exceptionally large value compared to most stars in the {\it Gaia} DR3 catalog \citep{2024A&A...688A...1C}. Various authors have shown that a RUWE$\gtrsim$1.4 is consistent with binarity \citep{2020MNRAS.496.1922B,2022A&A...657A...7K,2024A&A...688A...1C}. RUWE is sensitive to binaries with large deviations of the photocenter, orbital periods close to the DR3 time baseline of 34 months, and nearby systems \citep{2020MNRAS.496.1922B,2024NewAR..9801694E}. Indeed, \hd was one of 813,00 stars in the recent {\it Gaia} DR3 non-single star catalog whose photometric deviations from a single star model were found to be compatible 
with an unresolved companion \citep{2023AA...674A...9H}. 

In Table \ref{table:optical} we list the orbital elements for \hd provided by the Gaia Non-Single Star catalog \citep{2023AA...674A...9H}.
%Table \ref{table:optical} we list several of the Campbell binary orbital elements derived from the Thile-Innes elements for \hd 
For unresolved binaries the photometric wobble depends on both the mass B=M2/(M1+M2) and light ratios $\beta$=I2/(I1+I2) of the primary(1) and secondary(2) stars. The semi-major axis of the {\it photocentric orbit} (a$_p$) and the semi-major axis (a) of the binary orbit are related to each other by a$_p$=a\,(B$-\beta$) \citep{1975ARA&A..13..295V}. The measured value of a$_p$ drops to zero when the mass (and luminosity) of the binaries are either equal (i.e., M1=M2) or when the primary greatly exceeds the secondary (i.e., M1$\gg$M2) \citep{2024NewAR..9801694E,2024A&A...688A...1C}. We can use this dependence and Kelper's third law to constrain the properties of the unseen companion star. Assuming the primary is a K7V star (see \S\ref{sec:hispec}) we can compute the values of B and $\beta$ for less massive secondaries from 
\citet{2013ApJS..208....9P} and compare the predicted value of a$_p$ with the observed value. A secondary with a spectral type of either M0.5 or M5.5 satisfies {\it Gaia's} astrometic constraints.

\begin{deluxetable}{c|cc|c}[!ht]
\label{table:optical}
\tabletypesize{\small}
\tablecaption{Observed and Derived Stellar Parameters for \hdp}
\setlength{\tabcolsep}{0.03in}
\tablewidth{0pt}
\tablehead{
\colhead{Property} &
\colhead{Value} &
\colhead{error} &
\colhead{Reference}
}
\startdata
\hline
\hline
\multicolumn{4}{c}{{\it Gaia} Astrometry and Kinematics} \\
\hline
Parallax (mas) & 29.77 & 0.64 & Gaia DR3 \\
Distance (pc) & 33.84 & 0.80 & Gaia dist \\
Total Proper Motion (mas yr$^{-1}$) & 146.5 & 3.0 & Gaia DR3 \\
{Radial Velocity} (km s$^{-1}$) & -47.85 & 0.53 & Gaia DR3 \\
Gaia RUWE & 19.715 & \nodata & Gaia DR3 \\
\hline
\multicolumn{4}{c}{{\it Gaia} and 2MASS Photometry} \\
\hline
Gaia G (mag) & 9.937 & 0.003 & Gaia DR3 \\
Gaia BP (mag) & 10.718 & 0.003 & Gaia DR3 \\
Gaia RP (mag) & 9.076 & 0.005 & Gaia DR3 \\
Gaia BP-RP (mag) & 1.642 & 0.006 & Gaia DR3 \\
V (mag) & 10.459 & 0.030 & Gaia DR3 Conversion \\
R (mag) & 9.678 & 0.032 & Gaia DR3 Conversion \\
I (mag) & 8.938 & 0.038 & Gaia DR3 Conversion \\
J (mag) & 7.867 & 0.035 & 2MASS \\
H (mag) & 7.210 & 0.042 & 2MASS \\
K$_s$ (mag) & 7.053 & 0.036 & 2MASS \\
%Distance modulus (mag) & -2.631 & 0.009 & \nodata \\
%Gaia M$_G$ (mag) &  7.306 & 0.016811 & Gaia DR3 \\
%M$_V$ (mag) &  7.828 & \nodata & \nodata \\
%M$_R$ (mag) &  7.047 & \nodata & \nodata \\
%M$_I$ (mag) &  6.307 & \nodata & \nodata \\
\hline
\multicolumn{4}{c}{{\it Gaia} Orbital Solutions} \\
\hline
Period (day) & 1101 & 44 & Non-Single \\
Eccentricity & 0.60 & 0.02 & Non-Single \\
Inclination (deg) & 91 & 1 & Non-Single \\
Photometric Semi-Major Axis (AU) & 0.315 & 0.004 & Non-Single \\
\hline
\multicolumn{4}{c}{{\it Gaia} and TESS Derived Properties} \\
\hline
%\multirow{3}{*}{Spectral Type} & K & \nodata & Gaia DR3  \\
% & K6.5 & \nodata & PM Table using BP-RP \\
% & K6.5 & \nodata & PM Table using M$_G$ \\
\multirow{1}{*}{Effective Temperature (K)} & 4107 & 122 & TIC\\
 & 4196 & \nodata & Gaia parms \\
%& ??? & ??? & Kiman et al. 2024$^b$\\
\multirow{1}{*}{Surface Gravity [log g]} & 4.52 & 0.11 & TIC \\
%& 4.0 & ? & VOSA (re-check this) \\
& 4.61 & \nodata & Gaia parms \\
\multirow{1}{*}{Mass (M$_\odot$)}& 0.640 & 0.080 & TIC\\
& 0.700 & 0.005 & Gaia Parms\\
Radius (R$_\odot$) & 0.727 & 0.068 & TIC \\
%Rotation Period & \nodata & \nodata & \nodata & \nodata & \nodata \\
Luminosity (L$_\odot$) & 0.135 & 0.010 & TIC \\
Age (Gyr) & 9.376 & 0.25& Gaia parms \\
\hline
\hline
\enddata
%\tablenotetext{}{PM Table = \cite{PecautMamajekTable2013}, \url{https://www.pas.rochester.edu/~emamajek/EEM_dwarf_UBVIJHK_colors_Teff.txt}}
%\url{https://vizier.cds.unistra.fr/viz-bin/VizieR?-source=I/352}
\tablecomments{RUWE=re-normalized unit weight error, Gaia DR3=\citet{2021AA...649A...1G},
Gaia dist=\citet{2021AJ....161..147B},
2MASS=\citet{2006AJ....131.1163S},
TIC=\citet{2019AJ....158..138S},
Non-Single=\citet{2023AA...674A...9H},
Gaia parms=\citet{2023AA...669A.104K}.}
\end{deluxetable}

\subsubsection{High Resolution Spectroscopy}\label{sec:hispec}

\hd was observed on 2019 June 28 with the CHIRON high-resolution spectrograph \citep{2013PASP..125.1336T} on the SMARTS 1.5-m telescope at Cerro Tololo Inter-American Observatory (CTIO). This observation was part of the RKSTAR (RECONS\footnote{\url{www.recons.org}} K Stars) Survey, which is targeting all $\sim$5000 K dwarfs in the solar neighborhood \citep{2022AAS...24030515H}. The RKSTAR Survey observing and data reduction methods are described in detail elsewhere \citep[e.g.,][]{2022AJ....164..174H,2023AJ....166...63J}. In brief, a CHIRON spectrum was obtained with resolution R=80,000 with a integration time of 900 s, and processed through a customized pipeline \citep{2021AJ....162..176P}.

Four spectral line features were chosen for further analysis of the stellar parameters: H$\alpha$ at 6563 \AA\ and Ca II at 8542 \AA\ as tracers of chromospheric activity, Li I at 6708 \AA\ as an age indicator for K dwarfs, and the Na I doublet at 5890/5896 \AA\ as a tracer of surface gravity \citep[see][and references therein]{2022AJ....164..174H}. In particular, comparing the H$\alpha$ and Li II lines is useful for distinguishing between young or mature and active or inactive K dwarfs \citep{2024IAUGA..32P2962H,2024AAS...24325905C}. We note that the Ca II H and K lines commonly used for activity-rotation studies \citep{2006PASP..118..617R,2022ApJ...929...80B} lie outside CHIRON's spectral range of 4150-8800 \AA. 

\begin{table}[!ht]
    \caption{Spectroscopic Results and Derived Stellar Properties for \hdp}\label{tab:chiron}
    \centering
    \begin{tabular}{ccccccccc}
        \hline
        EW[H$\alpha$] & EW[Na I D] & EW[Ca II] & EW[Li I] & T$_{\rm eff}$ & R$_*$ & [Fe/H] & log\,$g$ & $v\,sin\,i$\\
        (\AA) & (\AA) & (\AA) & (\AA) & (K) & (R$_\odot$) & (dex) & (dex) & (km s$^{-1}$) \\
        \hline
        \hline
        0.52$\pm$0.004 & 5.75$\pm$0.62 & 0.68$\pm$0.23 & 0.0$\pm$0.0 & 4139$\pm$43 & 0.68 $\pm$ 0.10 & 0.13$\pm$0.02 & 4.66 & 8.68 $\pm$ 0.14\\
        \hline
    \end{tabular}
    \vspace{1cm}
\end{table}

The equivalent width (EW) measurements of these spectral features are summarized in Table \ref{tab:chiron} for \hdp. The H$\alpha$ line shows absorption with an EW of $0.52 \pm 0.004$ \AA, and the Ca II line at 8542 \AA\ has an EW of $0.68 \pm 0.23$ \AA. These values are consistent with those of a mature and chromospherically inactive K dwarf. If significant chromospheric activity were present, these lines would exhibit lower or even negative EWs, indicative of emission. The absence of such activity suggests a stable stellar environment. The Li I line at 6708 \AA\ was not detected.
%, with an EW of $0.0 \pm 0.0$ \AA. 
This non-detection is expected for older K dwarfs, as lithium is depleted in stars after a few hundred million years. In contrast, the Na I doublet at 5890/5896 \AA\ shows moderate absorption, with an EW of $5.75 \pm 0.62$ \AA, consistent with the surface gravity of a main-sequence mid-K dwarf. The combination of weak activity indicators and the absence of lithium supports the classification of \hd as a mature, quiescent star.

\begin{figure}[!ht]
\centering
\includegraphics[width=0.9\textwidth,angle=0,trim=0cm 0cm 0cm 0cm]{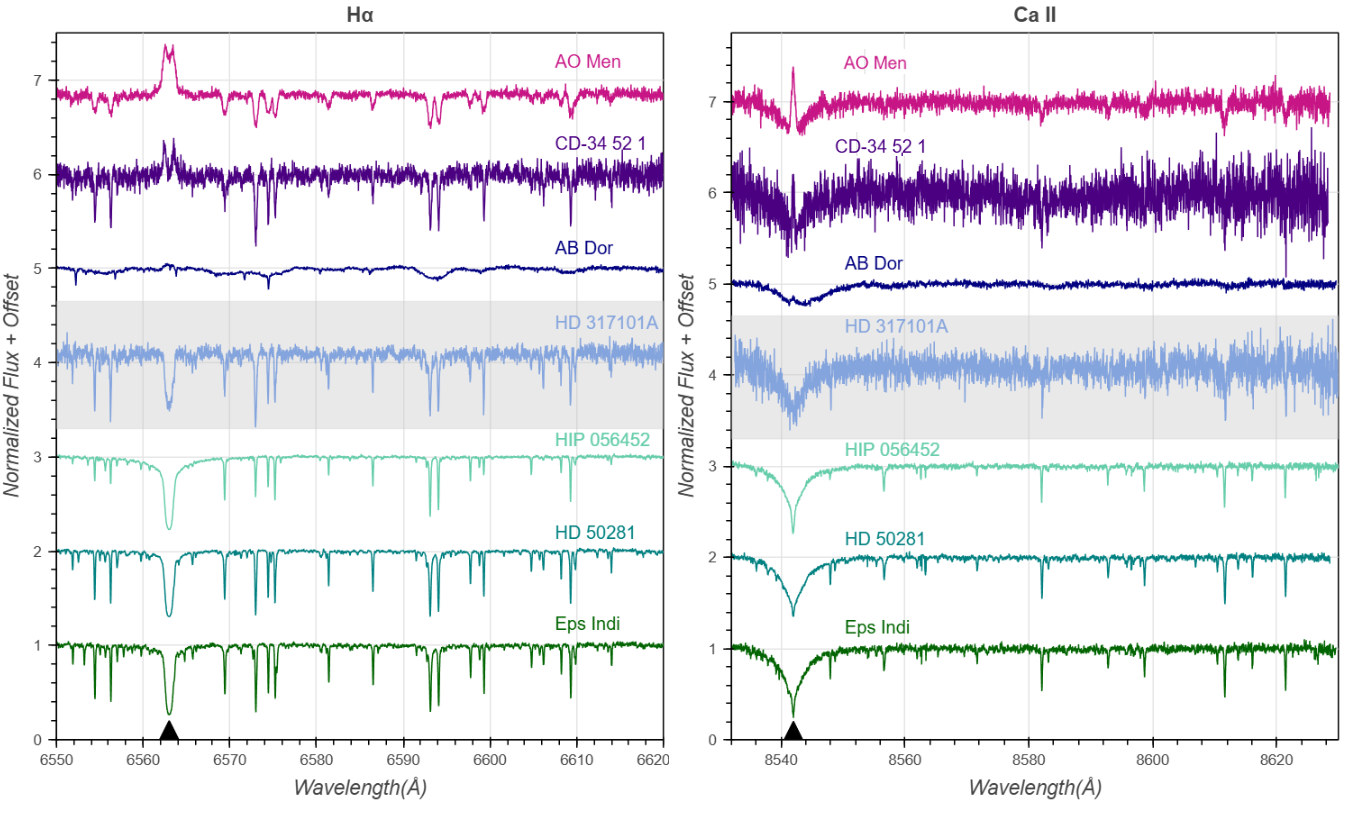}
\caption{Left: A compilation of stellar spectra for seven K dwarf stars ($B_p-K$ between 2.0 and 4.0) across multiple age groups, focusing on a window around the H$\alpha$ line at 6563 \AA\ indicated by a filled in black arrow. 
Right: A compilation of the same set of spectra focusing on a window around the Ca II IRT line at 8542 \AA\  indicated by a filled in black arrow. Comparison stars included from top to bottom: AO Men ($\beta$ Pic MG), 20 Myr, CD-34 52 1 (Tuc Hor MG, 40 Myr), AB Dor (AB Dor MG, 145 Myr), \hd (Radio Loud K Dwarf), HIP 056452 (field K Dwarf, $>1$\,Gyr), HD 50281 (Field K Dwarf $>1$\,Gyr), and $\epsilon$ Indi (field K Dwarf), $>1$\,Gyr). Spectral data normalized for comparison and offset for clarity.
}
\label{fig:optical_spectra}
%\vspace{1cm}
\end{figure}

Figure \ref{fig:optical_spectra} illustrates the H$\alpha$ (left panel) and Ca II IRT (right panel) lines of \hd compared to K dwarfs spanning a range of ages and activity levels. The stars above \hd in the plot represent young or active K dwarfs from the $\beta$ Pic Moving Group (20 Myr), Tucana-Horologium Association (40 Myr), and AB Dor Moving Group (145 Myr) \citep{2018ApJ...856...23G}. These young stars show significant chromospheric activity, with H$\alpha$ either fully in emission (e.g., AO Men in the $\beta$ Pic group) or filled in (e.g., CD-34 52 1 in the Tuc Hor group). Similarly, their Ca II profiles exhibit elevated activity levels consistent with strong chromospheric emission.

In contrast, the stars below \hd in Figure \ref{fig:optical_spectra} are mature field K dwarfs with ages exceeding 1~Gyr, including HIP 056452, HD 50281, and $\epsilon$~Indi. These older stars exhibit H$\alpha$ and Ca~II entirely in absorption, with no evidence of chromospheric activity. The similarity between \hd and these quiescent field stars highlights its inactive nature, distinguishing it from the younger, more active stars plotted above. Additionally, the spectra of the young stars - particularly AB~Dor - show clear signs of rotational broadening, indicative of rapid rotation, a common trait in stars younger than $\sim$150~Myr such as those in the $\beta$~Pic and Tuc~Hor moving groups. This broadening is not observed in \hd or the mature field K dwarfs, further supporting the interpretation of \hd as a slowly rotating, chromospherically inactive mid-K dwarf.
%Hodari Final Edit

We used a stellar library of 404 late-type stars to derive the stellar properties of \hd, including effective temperature (T$_{\rm eff}$), metallicity ([Fe/H]), surface gravity (log\,$g$), and stellar radius ($R_*$), using the Empirical SpecMatch code \citep{2017ApJ...836...77Y}. This code provides typical uncertainties of 70~K in T$_{\rm eff}$, 0.12~dex in [Fe/H], and 10\% in $R_*$. The derived values for \hd are listed in Table \ref{tab:chiron}: T$_{\rm eff} = 4139 \pm 43$~K, [Fe/H] = $0.13 \pm 0.02$~dex, log\,$g$ = 4.66~dex, and $R_* = 0.68 \pm 0.10\,R_\odot$. These properties are consistent with a mid-K dwarf (K7V), with a metallicity slightly above solar and a surface gravity and radius typical of a main-sequence star. The empirical radius estimate in particular aligns well with expectations for a star of this spectral type and effective temperature, further supporting our classification of \hd as a mature, mid-K dwarf.

We also estimated the projected rotational velocity (V$\sin{i}$) of \hd using a clean spectral region between 5440-5470\,\AA\ within the Empirical SpecMatch framework. The best-fit composite spectrum was constructed from five reference stars, and a weighted average of their known $v\,sin\,i$ values—using the fit coefficients as weights—yielded $v\,sin\,i$ = 8.68 $\pm 0.14$\,km\,s$^{-1}$. This moderate projected velocity is consistent with a slowly rotating K dwarf and provides additional support for its mature evolutionary state \citep{2022AJ....164..174H}.

We see no clear evidence of binarity in the high-resolution spectral lines, suggesting that the observed spectrum is primarily dominated by the light of the K dwarf. This allows us to reliably use the derived T$_{\rm eff}$ to estimate the star’s spectral type. The lack of significant chromospheric activity, combined with the derived metallicity and effective temperature, supports the interpretation of \hd as a relatively stable and mature star. This assessment aligns with the H$\alpha$ and Ca II measurements (Figure \ref{fig:optical_spectra}), which show absorption profiles consistent with a quiescent chromosphere, as well as with the low activity levels inferred from X-ray observations (\S\ref{sec:xrays}). Taken together, these results characterize \hd as an inactive mid-K dwarf with properties typical of a star that has evolved for at least several hundred million years, if not longer.

%Summary of what was done: 

%1) Line about rotational broadening added for AB Dor. 

%2) Radius estimate given for HD 317101A = 0.68 \pm 0.10\,R_\odot$ this value was derived from Emperical Specmaatch as well. Added to both the table and the text.

%3) Vsini estimate also found using the spectral library. Vsini for HD 317101A = 8.68 $\pm 0.14$\,km\,s$^{-1}$

%To dos (basically done)

%If there is some measure of rotation rate like V$\sin{i}$ that could be useful too. The X-ray data points to a rotation period of order 7-15 days.}

%Note: I see that Yee et al have a radius estimate. That was not in Hodari's email but it would be very useful to have that number.} 

%done

%\textcolor{red}{We used the stellar library of 404 late-type stars to derive the stellar properties of the star, including effective temperature (T$_{eff}$), the stellar radius (R$_*$), metallicity ([Fe/H]) and surface gravity (log g) \citep{2017ApJ...836...77Y} These authors estimate that their Empirical SpecMatch code has a typical accuracy of 70 K for T$_{eff}$, 10\% for R$_*$ and 0.12 dex in [Fe/H].

\subsubsection{Speckle Interferometry}\label{sec:speckle}

Higher resolution optical images are needed for two reasons. First, the 
position of the radio source (\S\ref{sec:meerkat}) is sufficiently uncertain that we need to verify that the radio emission was coming from \hd and not a background star in this crowded bulge field. Secondly, we hope to be able to detect the companion star, or at least constrain its properties. Thus, we carried out high spatial resolution observations on 2024 July 13 using the Zorro speckle interferometric instrument on Gemini South, with data collected simultaneously in two narrow-band filters centered on 562 nm and 832 nm \citep{2022FrASS...9.1163H}. Details of the instrument, data collection and reduction are found in \citet{2021FrASS...8..138S} and \citet{2011AJ....142...19H}. 

\begin{figure}
\centering
\includegraphics[width=0.48\textwidth,angle=0,trim=0cm 0cm 0cm 0cm]{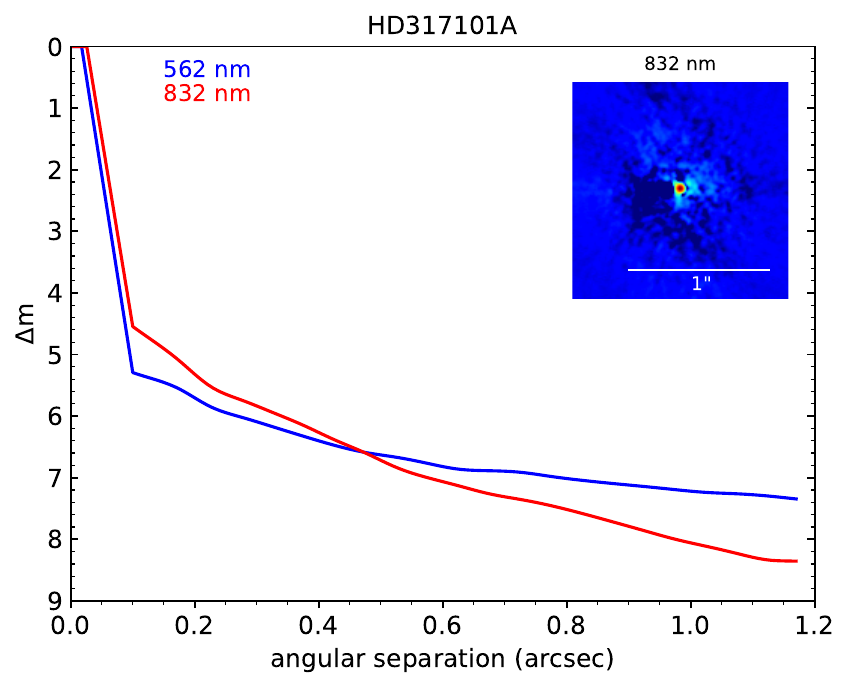}
\caption{Speckle image and 5-$\sigma$ contrast curves from the Zorro instrument on Gemini South. Blue camera (562 nm) and red camera (832 nm) curves are plotted. These curves show the magnitude difference ($\Delta$m) above which a field/companion star could be distinguished from the light of \hd with angular separations between 0.1$^{\prime\prime}$ and 1.2$^{\prime\prime}$.}
\label{fig:speckle}
\vspace{1cm}
\end{figure}

The output of this analysis are diffraction-limited images and contrast magnitudes ($\Delta$m) as a function of angular separation ($\Delta\theta$) from \hd at 562 nm and 832 nm (Fig.\,\ref{fig:speckle} and Table \ref{tab:speckle}). These limits agree with and improve upon speckle measurements made of \hd on 2022.2909 UT with the HRCam instrument on the 4.1-m SOAR telescope by \citet{2023AJ....165..180T}. 

%The output of this analysis are diffraction-limited images and contrast curves as a function of angular separation from the target star at 562 nm and 832 nm (Fig. \ref{fig:speckle}). The $5\sigma$\ contrast limits at 562 nm (832 nm) are $\Delta$m=5.14 (4.08), 6.64 (6.91) and 7.26 (8.10) at separations of 0.1$^{\prime\prime}$, 0.5$^{\prime\prime}$, and 1$^{\prime\prime}$, respectively. These limits agree with and improve upon speckle measurements made of \hd on 2022.2909 UT with the HRCam instrument on the 4.1 m SOAR telescope by \citet{2023AJ....165..180T}. No background or companion star was found at a resolution limit of 45 mas  in the I band with $\Delta$m=3.03 and 4.29, at separations of 0.15$^{\prime\prime}$ and 1$^{\prime\prime}$, respectively.

The larger offsets are useful for ruling out a field star as the source of the radio emission. Given the magnitude and colors of a star with the spectral type of \hd \citep{2013ApJS..208....9P}, these speckle observations would have detected field stars of spectral type M8V or earlier for offsets of 0.8$^{\prime\prime}$ or more.  As our radio position error is conservatively of this same magnitude, if the radio emission is not coming from \hdp, then it originates from a star of later spectral type and hence is fainter than the speckle limits.

Likewise, these contrasts can help identify the spectral type of the companion star. We earlier showed that the {\it Gaia} orbital solutions allow for a secondary whose spectral type is either M0.5 or M5.5. A companion star as bright as M0.5 would have easily been resolved at the resolution of these Zorro speckle images. Thus, the speckle data, used in combination with the {\it Gaia} astrometry constrain the secondary to be a M5.5 star 
%with mass 0.12 M$_\odot$ 
\citep{2013ApJS..208....9P}. Brown dwarf companions, while faint enough to remain undetected in the speckle images, would not be bright enough to produce the observed photometric wobble seen by {\it Gaia}.

\begin{table}[!ht]
 %   \caption{Speckle Contrast Limits at 562 and 832 nm versus Angular Distance from \hdp} \label{tab:speckle}
    \caption{Speckle Contrast Limits} \label{tab:speckle}
    \centering
    \begin{tabular}{lcc}
        \hline
           $\Delta\theta$ & $\Delta$m$_{562}$ & $\Delta$m$_{832}$ \\
           ($\arcsec$) & (mag.) & (mag.) \\
        \hline
        \hline
0.100 & 5.14 & 4.08 \\
0.175 & 5.61 & 5.47 \\
0.225 & 5.89 & 5.37 \\
0.275 & 6.01 & 5.85 \\
0.325 & 6.17 & 5.89 \\
0.375 & 6.33 & 6.13 \\
0.425 & 6.47 & 6.44 \\
0.475 & 6.63 & 6.52 \\
0.525 & 6.64 & 6.91 \\
0.575 & 6.77 & 6.99 \\
0.625 & 6.89 & 7.11 \\
0.675 & 6.89 & 7.32 \\
0.725 & 6.87 & 7.30 \\
0.775 & 7.01 & 7.49 \\
0.825 & 7.02 & 7.54 \\
0.875 & 7.11 & 7.76 \\
0.925 & 7.14 & 7.82 \\
0.975 & 7.19 & 8.04 \\
1.025 & 7.26 & 8.10 \\
1.075 & 7.24 & 8.22 \\
1.125 & 7.30 & 8.38 \\
1.175 & 7.35 & 8.36 \\
        \hline
    \end{tabular}
    \vspace{1cm}
\end{table}

\subsubsection{TESS Optical Light Curve}\label{sec:tess}

\begin{figure}
\centering
\includegraphics[width=0.55\textwidth,angle=0,trim=0cm 0cm 0cm 0cm]{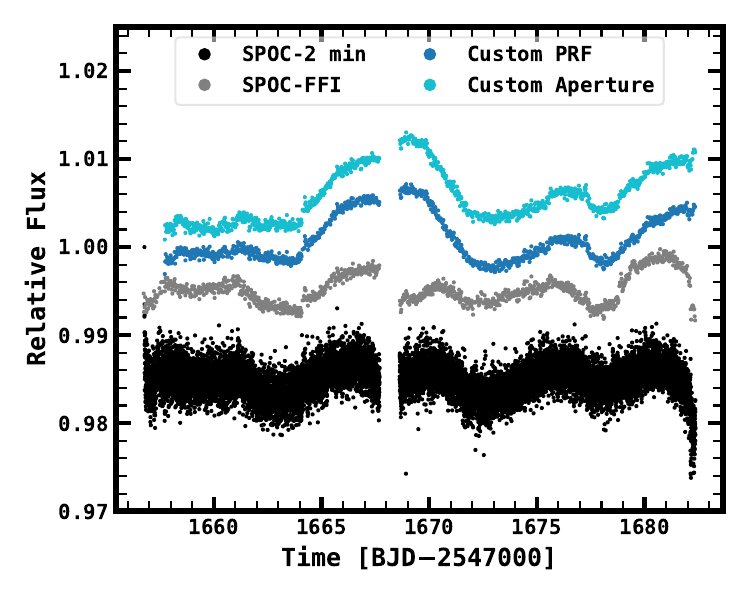}
\caption{Normalized TESS light curve of the star \hd (TIC\,407888149) observed in sector 13.  Four light curves are shown. From bottom to top these are (a) the SPOC 2-minute cadence, (b) the SPOC 30-minute cadence from the full frame images, and custom light curves extracted using {\it tess-phomo} from the full frame images using (c) point response function photometry, and (d) aperture photometry. See text for more details.}
\label{fig:tesslightcurve}
\vspace{1cm}
\end{figure}

To study the variability of \hd in the optical, we extracted a light curve of the system from the Transiting Exoplanet Survey Satellite \citep[TESS;][]{2015JATIS...1a4003R}. \hd was observed in sector 13 during the TESS primary mission, with a cadence of 120 seconds and also at a cadence of 30 minutes in the full frame images \citep{https://doi.org/10.17909/hzq3-6s38, https://doi.org/10.17909/f2t0-d696}. The TESS Science Processing Operations Center (SPOC) provided a light curve of \hd at both cadences. In principle the two light curves should have identical noise properties when adjusting for cadence, however slight differences in the detrending process often lead to differences in the light curves, particularly with regard to preserving low frequency (periods of several days and longer) variability. Because the SPOC light curves are designed primarily for the detection of transiting exoplanets, the default detrending algorithm, originally developed for the Kepler mission \citep{stumpe2014}, can sometimes suppress low-frequency variability caused by astrophysical sources such as rotational modulation \citep[e.g.,][]{ martinezpalomera2023, claytor2024}. Therefore, we also extracted custom light curves of \hd in the full frame images using both aperture and point response function (PRF) photometry via the python package {\it tess-phomo} \citep{wilson2025}. To preserve low-frequency variability, we de-trended the raw light curves by fitting a linear regression model that included instrumental model terms and a 1D spline with knots spaced 1 day apart and then removed the instrument model from the light curve. The four resulting light curves are shown in Fig.\,\ref{fig:tesslightcurve}.

\begin{figure}
\centering
\includegraphics[width=0.55\textwidth,angle=0,trim=0cm 0cm 0cm 0cm]{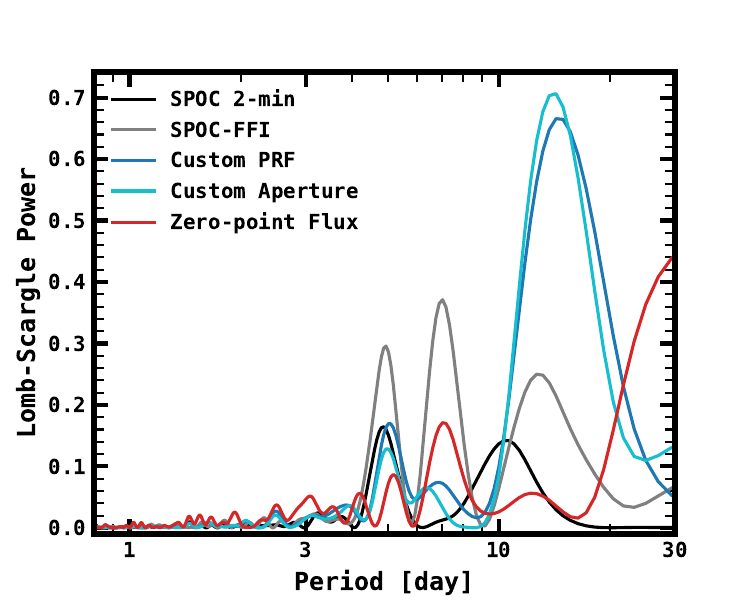}
\caption{Lomb-Scargle periodograms of four TESS light curves made toward the star HD 317101A (TIC 407888149). We argue that the periodicity around P$_{rot}\simeq$14 d is real and likely related to stellar rotation, while the peaks $\sim$5 d and $\sim$7 d are likely not real. The red curve shows a Lomb-Scargle periodogram of the TESS zero-point magnitude as modeled by {\it tessphomo}. The peaks seen in the zero-point flux near $\sim$5 d and $\sim$7 d are likely driven by either a background star or an instrumental artifact contaminating the light curves. See text for more details.}
\label{fig:tessperiod}
\vspace{1cm}
\end{figure}

 The photometry from TESS shows significant variability on timescales of a few days. To check for a rotation signal, we analyzed all four light curves for periodic signals using a Lomb-Scargle periodogram, as implemented in astropy \citep{2022ApJ...935..167A}. The resulting periodograms are shown in Fig.\,\ref{fig:tessperiod}. The two SPOC light curves show significant variability at periodicities of $\sim$5 d and $\sim$7 d, while our custom light curves show significant variability at a periodicity P$_{rot}\simeq$14 d, although given the duration of the observing sector this rotation period is highly uncertain. Of these peaks, we argue that the peaks in the periodogram at $\sim$5 d and $\sim$7 d are not real. {\it tess-phomo}, as part of the light curve extraction, models the TESS zero-point magnitude at each cadence by measuring the ensemble brightness in the background stars, which generally serves as a good proxy for systematic errors in the time-series photometry. Because the zero-point magnitude (red line in Fig.\,\ref{fig:tessperiod}) shows significant periodicity at $\sim$7 d and to a lesser extent $\sim$5 d, it is likely that these periodicities derive from either instrumental artifacts or a background star that is contaminating the light curve of \hdp. Thus, we conclude that there is likely some rotational modulation in the TESS light curve, but due to the crowded bulge field of \hd, we cannot definitively localize the variability to \hd itself. We also caution that the derived rotational period of $\sim$13-15 d is highly uncertain due to TESS’s short observing baseline ($\sim$25 days) and more data are needed to definitively measure a rotation period of this length, particularly from optical telescopes with a smaller pixel scale (compared to TESS’s pixel scale of $\sim{21}^{\prime\prime}$/pixel) and longer observing baseline. 

We also performed a transit search on the custom PRF light curve to determine if a transiting planetary companion is present. The light curve was flattened using the {\it wotan} package \citep{2019AJ....158..143H}, which is a comprehensive tool for removing out-of-transit variability in the light curve, primarily from stellar variability. We tested multiple different algorithms implemented in {\it wotan} for flattening the light curves and found consistent results. Using the flattened light curve we searched for a transiting planet using the transit least squares (TLS) algorithm \citep{2019A&A...623A..39H}. TLS is similar to the standard Box Least Squares (BLS) algorithm, except that it uses a realistic transit model that includes planetary ingress and egress. TLS is demonstrated to be more reliable than BLS, especially for the detection of small planets \citep{2019A&A...623A..39H}. 

We ran a TLS search on the flattened light curve. We searched for transits with orbital periods from 0.47 to 12.3 days. The minimum period is defined by the Roche Limit of the K star, which is calculated using the stellar mass and radius from Table \ref{table:optical}. The maximum period is defined as half the time span of the dataset, which ensures that at least transits would be visible in the data. The transit durations searched are defined by the default TLS settings. The search yielded no evidence of a planet transiting \hdp.

\section{Discussion}\label{sec:discuss}

\subsection{The Radio Emission Mechanisms}\label{sec:mechanisms}

The time-dependence of the stellar radio emission, its spectrum, brightness temperature and polarization properties can be useful diagnostics for identifying the underlying emission mechanism(s). At 1.5 GHz \hd shows activity on weekly and monthly timescales. Our best estimates of the duration of this emission comes from the 2020 MeerKAT monitoring in which the source is seen to be active on timescales of several hours (\S\ref{sec:meerkat}). At 3 GHz the source is only detected once over the several years of VLASS observations. While the duration ($\Delta$t) is not well constrained, the 3-GHz emission was at least seconds in duration, but it did not last beyond several minutes (\S\ref{sec:archive}). There does not appear to be any quiescent emission at a level of 40-80 $\mu$Jy at 1.5 GHz and 200 $\mu$Jy at 3 GHz (see Table \ref{tab:radio}).

The continuum spectra of \hdp, taken at 1.5 GHz and 3 GHz, also appear to be distinct from each other. The radio spectrum can be approximated by a simple power law S$_\nu$=S$_\circ\times\nu^\alpha$, where S$_\nu$=1--4 mJy and $\alpha\simeq{-4.4}$ at 1.5 GHz (\S\ref{sec:meerkat}), and S$_\nu$=1.2 mJy and $\alpha\simeq{0}$ at 3 GHz (\S\ref{sec:archive}). At 3 GHz there is a suggestion that the spectrum drops off at both the high and low frequencies. More comprehensive fits in \S\ref{sec:meerkat} show the 1.5-GHz data is consistent with a flat spectrum below 1 GHz and a steep decline $\alpha\simeq{-5.5}$ above. This unusual behavior is seen in the 2020 June and July MeerKAT epochs and it persists if we average all the data together, or just look at peaks. We attempted to confirm this behavior with the GMRT (\S\ref{sec:gmrt}) and the 2021 MeerKAT observations (\S\ref{sec:meerkat}). \hd is never as bright during these observations but a flat spectrum can be fit to the brightest MeerKAT epoch on 2021 May 07. Thus we cannot rule out two different slopes for the continuum spectra at 1.5 GHz.

The fractional bandwidth ($\Delta\nu/\nu$) is not well constrained. At 3 GHz, we detect emission across the entire 2-4 GHz band, with slight drop offs at high and low frequencies ($\Delta\nu/\nu\gtrsim{0.7}$). In contrast the 2020 MeerKAT spectra (Fig.\,\ref{fig:meerkat}) show the spectrum dropping sharply above 1 GHz. We attempted to better constrain the spectrum below 1 GHz using archival data. A continuation of the steep spectral index seen above 1 GHz to lower frequencies is not supported by the 150-MHz to 888-MHz limits (Table \ref{tab:radio}) provided that the source was active when these data were taken. The strongest evidence in support of a turnover are the 50 VAST-low non-detections taken between 2019 and 2024. Unlike at 3 GHz the 1.5-GHz emission appears to be a more narrow band phenomena ($\Delta\nu/\nu\lesssim{0.5}$). The estimates for the duty cycle ($\Delta{t}/t$) are likewise uncertain (see \S\ref{sec:archive}). However, we do have some evidence that the emission at 1.5 GHz is periodic and may be beamed (\S\ref{sec:period}), while at 3 GHz we do not have sufficient data to test for periodicity.  Finally, we note that the degree of circular polarization ($\vert$V/I$\vert$) was 90\% and $<27$\% at 1.5 GHz and 3 GHz, respectively (see \S\ref{sec:meerkat} and \S\ref{sec:meerkat}).

A brightness temperature (T$_b$) can be estimated from the standard equation \citep[see][]{2002ARA&A..40..217G}, which at the distance of \hd can be written as T$_b$=2.36$\times{10^{10}}$\,S$_\nu\times\nu^{-2}\times$\,r$_\odot^{-2}$, where the flux density is expressed in units of mJy, the observed frequency in GHz, and the emission radius is normalized to the solar radius (R$_\odot$). The size of the emission region is not well constrained but we use a range based on the photospheric radii of red dwarfs from K6 to M4 of approximately 0.7 to 0.35 R$_\odot$ , respectively \citep{PecautMamajekTable2013}. This is likely an overestimate of the length scale of the emission region, and thus a conservative estimate of T$_b$ \citep{2021A&A...648A..13C}. For the peak flux densities observed at 1.5 GHz and 3 GHz, we estimate T$_b$ values of 0.2-0.8$\times{10^{12}}$ K and 0.6-2.6$\times{10^{10}}$ K. The 1.5-GHz emission is close to the inverse-Compton limit of 10$^{11}-10^{12}$ K for incoherent synchrotron emission \citep{1994ApJ...426...51R}, while the 3-GHz emission is well below this value. For these same flux densities we can also estimate the spectral radio luminosity L$_r$, given by L$_\nu$=4$\pi$d$^2$\,S$_\nu$. In Table \ref{tab:emprop} we summarize the observed properties of the detected radio emission at 1.5 GHz and 3 GHz discussed above. 

If we compare the 1.5-GHz and 3-GHz properties in Table \ref{tab:emprop} phenomenologically there appears two types of transient radio behaviors, suggestive of at least two distinct emission mechanisms. At 3 GHz we see broad-band, time-variable milliJansky emission on timescales of seconds to minutes that is unpolarized and has a low brightness temperature. These emission properties are in line with {but do not conclusively prove a gyro-synchrotron emission origin \citep{2002ARA&A..40..217G}. In contrast, the properties of the 1.5-GHz emission (T$_b$, V/I, $\alpha$, duration and periodicity) support the hypothesis that the emission from \hd is coherent and, given the high circular polarization, is likely produced by the electron cyclotron maser instability process \citep[ECM;][]{2002ARA&A..40..217G}. The sharp spectral cut off is predicted from ECM emission and it implies a gyro-frequency $\nu_c\simeq$1.5 GHz, from which we derive a magnetic field strength from Eqn.\,\ref{eqn:ecm} of B=0.36 kG.

\begin{table}
    \caption{Radio Emission Properties}\label{tab:emprop}
    \centering
    \begin{tabular}{rrr}
        \hline
        \hline
        $\nu$ (GHz) & 1.5 GHz & 3 GHz \\
        S$_\nu$ (mJy) & 0.2-4 & 1.2 \\
        L$_r$ (erg s$^{-1}$ Hz$^{-1}$) & 1.4-5.5$\times{10^{15}}$ &1.6$\times{10^{15}}$ \\
        $\Delta\nu/\nu$ & $\lesssim{0.5}$ & $\gtrsim{0.7}$ \\
        $\Delta{t}/t$ (\%) & $\sim{25}$ & $\sim{17}$\\
        $\alpha$ & $-5.5$  & 0 \\
        T$_b$ (K) & 5$\times{10^{11}}$ & $10^{10}$ \\
        $\vert${V/I}$\vert$ (\%) & 90 & $<27$ \\
        $\Delta$t & hrs-min & sec-min \\
        P (days) & 3.72 & \nodata \\
        \hline
    \end{tabular}
    \vspace{1cm}
\end{table}

\subsection{Chromospheric Activity on K dwarfs}\label{dis:kstar}

Could an active chromosphere in the K star primary of \hd be the source of the radio emission summarized in Table \ref{tab:emprop}? K dwarfs can give rise to time-variable and/or steady radio emission \citep{2002ARA&A..40..217G,2019PASP..131a6001M}. All of the K stars detected to date exhibit signs of enhanced chromospheric activity, including bright UV or soft X-ray emission, rapid rotation, H$\alpha$ or Ca II, H, and K lines in emission \citep{1983ApJ...274..776L,1992A&A...264L..31G}. Binarity seems to play an important role since all existing samples of radio-active K dwarfs appear to be close binaries or recent binary mergers, including tidally locked systems such as RS Canum Venaticorum binaries (RS CVn), FK Comae stars and BY Draconis systems \citep{2021A&A...654A..21T,2023PASA...40...36D,2024arXiv240407418D,2024A&A...684A...3Y,2024arXiv240407418D}. The AB Doradus system could possibly be an analog to \hdp. Both systems are visual binaries with the radio activity coming from AB Dor A, a young, fast rotating (P$_{rot}$=0.514 d) K1V star with a close companion (AB Dor C) of spectral type M6$\pm$1 \citep{1994ApJ...430..332L,2006ApJ...638..887L,2017A&A...607A..10A}. The range of radio luminosities (log\,L$_r$=14.7-15.3) and the brightness temperatures (log\,T$_b$=8.7-9.4) from AB Dor A are comparable to the 3-GHz values we measure in Table \ref{tab:emprop}.

The emission from K stars at GHz frequencies is thought to arise predominately from non-thermal gyro-synchrotron radiation generated from mildly relativistic electrons accelerated by flares during magnetic reconnection events in the corona \citep{1985ARA&A..23..169D}. Strong support for interpreting this emission as incoherent gyro-synchrotron comes from brightness temperature measurements, and the observed spectral and polarization properties of radio K stars \citep{1985ApJ...289..262M,1987AJ.....93.1220M}. Further evidence for this interpretation can be found in a tight correlation between the radio spectral luminosity (at 5 GHz) and the soft X-ray luminosity, spanning 10 orders of magnitude, from solar flares to the brightest RS CVn binaries \citep{1994A&A...285..621B}. The underlying physical mechanism powering this empirical relationship posits that the same energy release that accelerates the non-thermal electrons heats the coronal plasma, giving rise to thermal X/UV Bremsstrahlung emission \citep{2010ARA&A..48..241B, 2022ApJ...926L..30V}. For those magnetically active systems that contain K stars, the X-ray luminosity is typically in the range of 5$\times{10}^{29}$ erg s$^{-1}$ to 10$^{32}$ erg s$^{-1}$, with radio spectral luminosities (at 5 GHz) of 3$\times$10$^{14}$ erg s$^{-1}$ Hz$^{-1}$ to 10$^{18}$ erg s$^{-1}$ Hz$^{-1}$.

Coherent radio emission has been detected from some late-type K stars that are found as a main sequence companion within short-period (P=1-30$^d$) interacting binaries known as RS CVn stars \citep{2021A&A...654A..21T}. At MHz frequencies these stars have high brightness temperatures (T$_b\gg10^{11}$ K) with circular polarization fractions of 50-90\%, and their emission has been interpreted as originating from the ECM process \citep{2022ApJ...926L..30V}. Somewhat surprisingly, these coherent sources (at 144 MHz) appear to follow the {G{\"u}del}--Benz relationship established for incoherent gyro-synchrotron sources. It is not well understood why the coherent emission from these chromospherically-active systems obeys this correlation. However, we note that they dominate the top end of the observed distribution, with typical radio spectral luminosities (at 144 MHz) of the order 10$^{16}$ erg s$^{-1}$ Hz$^{-1}$ and X-ray luminosities  $\sim$few$\times{10}^{31}$ erg s$^{-1}$ \citep{2022ApJ...926L..30V}.

The optical properties of \hd are unlike the K stars discussed above. The optical spectrum (\S\ref{sec:hispec}) and light curve (\S\ref{sec:tess}) all indicate that \hd is an older, quiescent dwarf. We find no support for the hypothesis that the radio emission from \hd originates from an active, flaring chromosphere of a K star. 

Further support for this quiescent K star hypothesis comes from the X-ray data. The X-ray luminosity (L$_x$) is a sensitive measure of the efficiency of the stellar dynamo in heating the coronal plasma \citep{2024LRSP...21....1K}. The average X-ray luminosity of 2.52$\times{10}^{28}$ erg s$^{-1}$ (\S\ref{sec:xrays}) lies toward the faint end of the luminosity distribution for quiescent X-ray emission from flare stars but is consistent with a volume-limited sample of K stars \citep{2015A&A...581A..28P,2004A&A...417..651S,2025arXiv250107313Z}. This value of L$_x$ is approximately 3 orders of magnitude below BY Draconis systems and other close binaries \citep{1992A&A...264L..31G}. Moreover L$_x$ is 2 orders of magnitude below most isolated radio-emitting K dwarfs, such as AB Dor.\,\citep{2024MNRAS.530.2442B}. AB Dor exhibits both short term X-ray flares and long term variations, neither of which is seen in the X-ray light curves of \hd \citep{2013A&A...559A.119L,2024MNRAS.527.1705D,2024ApJ...966...86S}

We can further quantify the level of activity for \hd since, for a given mass and radius, there is an empirical relationship between the rotation rate of the star and its X-ray luminosity. This is best expressed using the R$_x$-R$_\circ$ relation, where 
R$_x$ is the X-ray luminosity of the star normalized by its bolometric luminosity, and R$_\circ$ is the Rossby number, expressed as the ratio of the rotation period (P$_{rot}$) to the convective turnover time ($\tau$) \citep{2011ApJ...743...48W, 2018MNRAS.479.2351W, 2021AA...649A..96J, 2022ApJ...929...80B}. Given our observed L$_x$ and the value of L$_{bol}$ in Table \ref{table:optical}, we estimate a Rossby number of 0.45 \citep{2011ApJ...743...48W}. This places \hd in the unsaturated regime of this activity-rotation relation. Radio-bright, rapidly rotating K stars are typically in the saturated regime such that R$_x$ is constant, independent of R$_\circ$. Given the mass of the primary star (Table \ref{table:optical}) and the relationship from \citet{2011ApJ...743...48W} we derive log$\,\tau$=1.43, or a P$_{rot}=10.6^d$. While this value of P$_{rot}$ is uncertain by at least 25\%, it is consistent with a median of 15.4$^d$ for large sample $v\,sin{i}$ measurements of late-type main sequence stars \citep{2013A&A...557L..10N}. We also note its similarity with the modulations in the TESS light curves (\S\ref{sec:tess}), for which we derived P$_{rot}\simeq$14 d. The slow rotation and the apparent lack of any X-ray flare activity (\S\ref{sec:xrays}) supports our claim that this is a quiescent K star.
%Prove this satuarted assertion with a plot of a few known radio bright stars

We might also expect that a chromospherically active K star would follow the {G{\"u}del}--Benz relationship \citep{1994A&A...285..621B}. This is decidedly not the case. Using L$_x$ and the best-fit to the relation from \citet{2014ApJ...785....9W}, we derive a value for L$_r$ that is $\sim$245 times (or 2.4 dex) less luminous than the radio emission that we observe. Some caveats apply as the radio and X-ray data were not taken simultaneously and, strictly speaking, the relationship applies only at 5 GHz. Moreover, while the K star has an M star companion, the orbital period of 1100 days is much larger than radio-loud interacting binaries that populate the top end of the {G{\"u}del}--Benz relationship, such the RS CVn systems discussed earlier.

Summarizing, the multi-wavelength properties of \hd argue against the hypothesis that its unusual radio emission originates from a chromospherically active K dwarf primary. Unlike typical radio-loud K stars, which are often found in close or tidally locked binary systems with strong magnetic activity, rapid rotation, and intense X-ray emission, \hd lacks all such signatures. Its optical spectrum and TESS light curves indicate a slowly rotating, older, and quiescent star. The measured X-ray luminosity lies well below levels typical of active binaries or isolated flaring K dwarfs, and shows no evidence of variability. Furthermore, the inferred rotation period places \hd in the unsaturated regime of the rotation-activity relation, inconsistent with the behavior of radio-bright, rapidly rotating K stars. Finally, the observed radio luminosity exceeds expectations from the {G{\"u}del}--Benz relation by over two orders of magnitude. These findings collectively rule out a magnetically active origin for the radio emission from the K dwarf primary.

With no evidence of an active corona or chromosphere in the primary star, we must look elsewhere to explain why this otherwise quiescent dwarf is radio loud. In the next two subsections, we examine whether the observed radio emission can instead be attributed to the M dwarf companion or a stellar planet interaction (SPI).

%WHAT MORE CAN WE SAY ABOUT THE X-ray proprties relative to a sample of K stars. and the bolometric luminosity (L$_{bol}$) for HD\,317101A from Table XX, 
%, and normalized value (L$_x$/L$_{bol}$) of 4.8$\times{10}^{-5}$. From the  stellar radius (r$_\star$) in Table XX, it follows that the surface X-ray flux (F$_x$=L$_x$/4$\pi$r$_{\star}^{2}$) is 7.8$\times{10^{5}}$ erg s$^{-1}$ cm$^{-2}$, from which we can infer a coronal temperature (T$_c$) of 3.7$\times{10^{6}}$ K \citep{2015A&A...578A.129J}. 

%What fraction of K stars are active?. The measured X and r are omconistsnet with an actice K star. No RS Cvn ID, the emission is coherent. Are there any isolated K stars that have (coherent) radio?

%"particular emission was recognized as being coherent, there has been an unresolved controversy as to whether the underlying mechanism is plasma or ECM emission." MHz obs and UV Ceti have settled it to be ECM.

%X-ray emission is a sensitive tracer for magnetic activity 

%If the variability of J180526.43$-$292953.2 is occurring on the scale of a stellar diameter (10$^{11}$ cm) or below then the brightness temperature exceeds 10$^{12}$ K, suggesting a coherent mechanism. 

%Understand the statement that M stars have properties of Sun-like stars and Jupiter-like stars. Also understand how to bst use the TESS data to characterize the flare rate of the stars(s) that TESS sees in taht light curve.

\subsection{Magnetic Activity From An Unseen Companion}\label{dis:mstar}

Given that the K star primary has low stellar activity (\S\ref{dis:kstar}), could the radio emission from \hd actually originate from a companion star? In this respect, \hd could resemble HD\,220242, a F5V star recently identified as a radio-luminous source in a Stokes V version of the LOFAR Two-Metre Sky Survey \citep[V-LoTSS;][]{2023A&A...670A.124C}. \citet{2024A&A...684A...3Y} argue that a UCD companion is the origin of the anomalous radio emission and not the F primary. In \S\ref{sec:optical} we showed that {\it Gaia} astrometric residuals revealed the presence of an unseen companion around the K primary star with an orbital period of 1100 days (\S\ref{sec:optical}). We used speckle observations (\S\ref{sec:speckle}) to further constrain the spectral type of the companion to M5.5, with a semi-major axis of 1.9 AU. Brighter stellar companions would have been detected in the speckle images, while fainter UCDs, including brown dwarfs, could not have produced the photometric wobble observed from {\it Gaia}. This is an interesting transition zone; stars with spectral types later than about M4 are fully convective \citep{2024ARA&A..62..593H}, but they are not cool enough to be members of the UCD population, defined to be M7 spectral types or later \citep{1997AJ....113.1421K}. The prototypical flare star UV Ceti (spectral type M6) also lies in this transition zone \citep{2024LRSP...21....1K}.

At radio wavelengths M stars show a diverse range of phenomenology, including quiescent or near steady-state emission, flares, and beamed, periodic emission \citep[e.g.,][]{2022ApJ...932...21K}. Extended monitoring of M flare stars show both short-duration, incoherent emission \citep{2019MNRAS.488..559Z,2021A&A...648A..13C,2022ApJ...935...99B,2024A&A...682A.170B,2024MNRAS.531..919Z}, and long-duration, coherent bursts \citep{2019ApJ...871..214V}. At GHz frequencies solar-like processes have been identified, including gyro-synchrotron emission which obeys the {G{\"u}del}--Benz relation.  At MHz frequencies \citet{2021NatAs...5.1233C} have shown that M stars across all spectral types and magnetic activity levels show hr-long duration ECM emission. \citet{2024A&A...684A...3Y} have recently increased the sample of M dwarfs by searching for {\it Gaia} matches with VLASS, LoTSS, and V-LoTSS. Independent of spectral type, they argue that these M stars emit coherent emission and that all of them deviate from the {G{\"u}del}--Benz relationship in the same manner (i.e., radio loud/X-ray quiet). 

The radio and X-ray properties of \hd share many similarities to the population of M stars. We have identified both long-duration, coherent ECM emission (at 1.5 GHz) that may be periodic, and short-duration, and possibly incoherent gyro-synchrotron emission (at least at 3 GHz). This dual behavior is also seen toward UV Ceti, with the ECM emission beamed at the 5.45 hr rotation period of the star \citep{2019MNRAS.488..559Z,2024ApJ...970...56P}. Likewise, the steady-state radio (and X-ray) luminosities of UV Ceti are similar to \hd \citep{2021AA...649A..96J,2022ApJ...935...99B} and the M dwarfs $\delta$ And B and GJ\,3237 from the survey of \citet{2024A&A...684A...3Y}. There appears to be a significant change in the radio/X-ray properties of M stars below M4 \citep{2014ApJ...785....9W}, with the X-ray luminosity dropping precipitously with an increasing number of radio-overluminous stars relative to the {G{\"u}del}--Benz relationship. It has been posited the dominant emission processes is undergoing a transition from chromospheric to auroral between M4 and M7 spectral types \citep{2017ApJ...846...75P,2024A&A...684A...3Y}.

The properties of the M star secondary of \hd differs in at least one important respect from many radio-active M stars in the apparent absence of flare activity. Late M4-M6 dwarfs show the highest level of optical flaring, with more than 40\% of them showing strong flares, for which we estimate from the sample of \citet{2020AJ....159...60G} a median bolometric energy E$_{f}=3\times$10$^{32}$ ergs and average duration of $\delta{t}$=360 s. \citet{2021ApJ...919L..10P} showed how estimates of the stellar flare rates from TESS could be used to identify which of the radio-emitting M stars with ECM emission were driven by a coronal process versus a possible SPI origin. We visually inspected the X-ray (\S\ref{sec:xrays}) and TESS 2-minute cadence (\S\Ref{sec:tess}) light curves for short-term variability and we see no evidence of flaring activity for a TESS flare rate of $<$0.04 d$^{-1}$. If the radio emission originates from the M star, the absence of flares suggests that the ECM emission may be from auroral magnetospheric activity and is not powered by coronal flares. However, the normalized TESS lightcurve from which we search for the M star flares is likely dominated by the light of the K star of \hd and possibly also HD\,317101B, 4.8$^{\prime\prime}$ away. Using expressions for the fractional amplitude for a given flare energy and duration from \citet{2020AJ....159...60G}, we estimate that a median flare whose energy and duration are given above would be masked by the light of the K star(s). However, the distribution of late M star flares energies is sufficiently broad \citep{2020AJ....159...60G} that 10x brighter flares  could have easily been detected in the TESS lightcurve (Fig.\,\ref{fig:tesslightcurve}). There is a similar factor of 100x difference in the luminosities of the M1+M6 binary GJ\,412 AB, and yet TESS is able to detect extensive flare activity from the radio-active secondary WX UMa \citep{2021ApJ...919L..10P}. Likewise, optical and X-ray flares with the frequency and magnitude of those from the prototypical flare star UV Ceti would have been detected toward \hd 
\citep{1992AJ....104.1161E,2003ApJ...589..983A,2004A&ARv..12...71G,2016A&A...589A..48S}. The absence of TESS flares is suggestive of a quiescent M star but it remains possible that numerous smaller flares, hidden by the light of the K star(s), could be driving an active chromosphere.

% X-rays are saturated regeime so we can't constrain the rotation period.

%It lies below the ``gap'' in which the $\alpha\Omega$ dynamo (operating at tachocline between the inner radiative and outer convective zones) powers the stellar B field \citep{2024ARA&A..62..593H}, 

%but lies above the UCD population, defined to be M7 spectral types or later \citep{1997AJ....113.1421K}. 

%A brown dwarf companion for \source seems unlikely. The observed L$_r$ and L$_x$ of \source are well in excess of the typical values of L and T dwarf luminosities at \citep{2014ApJ...785....9W,2017ApJ...846...75P,2018ApJS..237...25K,2022ApJ...932...21K,2024MNRAS.tmp..938K}. However, this hypothesis cannot be ruled out since the T dwarfs BDR\,J1750+3809 and WISE J062309.94$-$045624.6 have radio spectral luminosities at or above $10^{15}$ erg s$^{-1}$ Hz$^{-1}$ at 144 MHz and 1.36 GHz, respectively \citep{2020ApJ...903L..33V,2023ApJ...951L..43R}.

\subsection{Star-Planet Interaction}\label{dis:spi}

Could the ECM emission originate from a star-planet interaction (SPI)? Specifically, the SPI in this instance would be from a planet inducing auroral currents on to the polar regions of a star. As we noted in  \S\ref{sec:intro}, a number of SPI claims of have been made, but none have been definitive. Several criteria need to be met (e.g., quiescence, periodicity and stability) to distinguish ECM from auroral versus stellar processes. Answering this question for \hd is complicated by the fact that we do not know if the ECM originates from the primary or secondary star. 

If the ECM emission is coming from the K star, then an SPI model is a plausible explanation. Although there remains some debate \citep{2022ApJ...929..169R}, K stars have been posited as promising ``Goldilocks'' targets to host habitable planets \citep{2016ApJ...827...79C}. At least one K star with an exoplanet has been seen undergoing polarized short-term radio flaring \citep{2018ApJ...857..133B}. Based on the optical spectrum and the X-ray and optical light curves, we showed in \S\ref{dis:kstar} that the primary of \hd is an old, slowly rotating star. Thus it is unlikely to give rise to strong multi-wavelength flares as seen in other radio-active stars. The K star is quiescent; a magnetically active chromosphere cannot explain the radio emission that we observe.

Instead, the properties of the ECM emission at 1.5 GHz more closely resembles that of the kHz to MHz auroral emissions from the magnetized planets in our solar system \citep{1998JGR...10320159Z,2021A&A...648A..13C}. As we discuss in \S\ref{sec:meerkat}, one of the more unusual properties of the radio emission is the apparent constancy of its spectral and polarization properties.  \citet{2019ApJ...871..214V} and \citet{2021A&A...648A..13C} suggested that the time-averaged ECM emission from SPI would be more stable than ECM produced by stochastic flare activity. Similar to the solar system planets, our time-averaged radio spectra (Fig.\,\ref{fig:meerkat}) also exhibit the same sharp drop in emission above the gyro-frequency ($\nu_c$). Although this is a fundamental spectral property of ECM emission \citep{2002ARA&A..40..217G}, it has rarely been measured in the time-averaged or dynamic spectra of other M stars and brown dwarfs \citep[but see][]{2023NatAs...7..569P}. It suggests that the ECM originates from a stable, large-sale magnetospheric configuration rather than a rapidly changing magneto-ionic environment, as expected from coronal loops. 

%From Eqn. 1 it follows that the polar magnetic field from which the auroral emission originates is XX kG. This is a reasonable value of a B field for a K star (REF). IS it???? Another point in favor of the auroral ECM model is the constancy of the spectral and polarization properties (handiness and degree). 

Making a strong case for SPI emission originating from the M star secondary of \hd is more complex.  We have argued in \S\ref{dis:mstar} that the lack of optical and X-ray flares suggests that the M star is quiescent, and points to an auroral ECM origin. This is not a strong argument. While ECM emission is ubiquitous among radio-emitting M stars and brown dwarfs \citep{2006ApJ...653..690H,2008ApJ...684..644H}, long duration ECM emission seems to be present in samples of M stars whether the chromosphere is quiescent or not \citep{2019ApJ...871..214V, 2021NatAs...5.1233C}. The stellar rotation rate could be another useful predictor but the same  R$_x$-R$_\circ$ analysis carried out for the K star in \S\ref{dis:kstar} implies that the M star is in the saturated regime where P$_{rot}$ is not well constrained \citep{2018MNRAS.479.2351W}. The single best argument in favor of auroral emission is the constancy of the spectral and polarization properties, suggestive of a stable magnetophere. However, unlike in the case of the K star primary, auroral emission is common among UCDs, including L and T dwarfs \citep{2014ApJ...785....9W,2017ApJ...846...75P,2018ApJS..237...25K,2022ApJ...932...21K,2024MNRAS.tmp..938K} and another source other than SPI could explain the emission (e.g., co-rotation breakdown).

%it leaves open the larger question of the source of the magnetoplasma engine powering this emission.

In principle, the strength of the B field, inferred from the sharp spectral break at 1 GHz (\S\ref{sec:mechanisms}), might allow us to identify which star is the source of the ECM emission. Late M stars have dipolar B fields measured as high as several kilo-Gauss \citep{2021A&ARv..29....1K}. While not as well studied as the more numerous M stars, late K dwarfs can also have significant magnetic fields, but their dipolar values are generally weaker than those found in fully convective M dwarfs \citep{2012LRSP....9....1R}. Field strengths of tens to several hundred Gauss are more typical, especially for older, slowly rotating K dwarfs \citep{2014MNRAS.441.2361V}. Stronger magnetic fields in K stars tend to be measured from younger, faster-rotating stars with corresponding higher levels of chromospheric and coronal emission \citep[e.g. AB Dor.;][]{1999MNRAS.302..437D}. Our measured magnetic field of B=0.36 kG lies in an intermediate zone; somewhat low for typical M stars and (perhaps) on the high end for old, slowly rotating K stars. 

% lots of UCBs have this emission. we are circumvent/evading/avoiding shift the question or re-frame the question into another more perplexing one, not specific to \hd but to the larger pop of UCDs.

%A brown dwarf companion for \source seems unlikely. The observed L$_r$ and L$_x$ of \source are well in excess of the typical values of L and T dwarf luminosities at \citep{2014ApJ...785....9W,2017ApJ...846...75P,2018ApJS..237...25K,2022ApJ...932...21K,2024MNRAS.tmp..938K}. However, this hypothesis cannot be ruled out since the T dwarfs BDR\,J1750+3809 and WISE J062309.94$-$045624.6 have radio spectral luminosities at or above $10^{15}$ erg s$^{-1}$ Hz$^{-1}$ at 144 MHz and 1.36 GHz, respectively \citep{2020ApJ...903L..33V,2023ApJ...951L..43R}.

The primary weakness of the SPI model is the same regardless of which star is implicated. Although the TESS time-series was not extensive, it showed no evidence for a planetary companion (\S\ref{sec:tess}). However, the true planet occurrence rate for K stars is in excess of 50\%, while transient instruments like TESS are sensitive to only a small fraction ($\sim$1\%) of planets \citep{2025AJ....169..112S}. There is preliminary evidence that the 1.5-GHz emission is periodic (\S\ref{sec:period}). The observed radio period of 3.7 days is intriguing. It is in the range of orbital periods expected in which a near-in planet can induce sub-Alfv\'enic interaction with a star's B field \citep{2020NatAs...4..577V, 2023NatAs...7..569P}. By analogy with the Jupiter-Io system, if SPI is responsible for the 1.5-GHz emission, it is expected that the radiation would be beamed toward the observer at distinct phases, with a synoptic period related to both the rotation of the star and the orbital period of the planet \citep{2011A&A...531A..29H,2021MNRAS.504.1511K, 2023MNRAS.524.3020K}. 

The evidence for periodicity or beaming is not strong enough to unambiguously claim of SPI. Establishing periodicity has been equally difficult in other putative SPI candidates. Periodic ECM is seen towards many UCDs, but none to date have been associated with the orbit of a known planet \citep[e.g.,][]{2022AJ....163...15C}. In the case of the M star YZ Ceti, the emission is seen at distinct phases of the planetary orbit, but flaring activity for the source of the emission cannot be fully ruled out \citep{2023NatAs...7..569P}.  If it could be firmly established that beamed, periodic ECM emission with P$\simeq$3.7 days originates from the K star, then an SPI origin seems likely. If instead, the emission originates from the M star, it remains possible that the periodic activity we see is modulated by the rotation period of the star \citep[e.g.,][]{2019MNRAS.488..559Z}. Unlike the K star for which optical and X-ray data measure P$_{rot}\simeq{10-15}$ days, we have no useful  P$_{rot}$ constraints for the M star. 

%Another weakness is the absence of a strong signature of beamed, periodic emission.  Although our constraints on the duty cycle of the 1.5 GHz radio emission from \hd (Table \ref{tab:emprop}) are consistent with beamed ECM emission, we do not have strong evidence for periodicity.  Future improved optical or radio light curves of \hd could bolster the case for a SPI.

\subsection{An Ultra Long Period Transient?}

With its strong variability, an active phase that persists for weeks or months, a high degree of (circular) polarization and a very steep radio spectrum, \hd shares some of the properties of other Galactic transients. The first of these objects to be found were the three Galactic center radio transients \citep[GCRT][]{2002AJ....123.1497H,2005Natur.434...50H,2007ApJ...660L.121H, 2009ApJ...696..280H}. The best studied of these is GCRT\,J1745$-$3009 which was seen to emit pulses with a period of 77 min, and is now thought to be a member of a new class of ultra-long period transients (ULPT). 

Some of the more recent ULPTs include GLEAM-X\,J162759.5$-$523504.3 with a period (P) of 18.18 minute \citep{2022Natur.601..526H}, GPM\,J1839$-$10 with P=22 minute \citep{2023Natur.619..487H}, ASKAP J193505.1+214841.0 \citep{2024NatAs.tmp..107C} with P=54 min, CHIME J0630+25 with P=421 s \citep{2024arXiv240707480D}, and possibly others \citep{2021ApJ...920...45W,2024arXiv240612352D,2024arXiv241115739L}.  A current list of ULPT is given in \citet{2025arXiv250103315R}. Among this heterogeneous sample GCRT\,J1745$-$3009 has a similar steep spectral index as \hdp, and the fractional circular polarization of GCRT\,J1745$-$3009 and ASKAP J193505.1+214841.0 during certain phases are comparable to \hd \citep{2010ApJ...712L...5R,2024NatAs.tmp..107C}. Similar to \hdp, several of these ULPTs also have optical/IR companions \citep{2024ApJ...976L..21H, 2024arXiv240811536D,2024NatAs.tmp..107C}. Collectively, however, the radio properties of the ULPTs are more similar to degenerate stars. This interpretation is supported by the recent spectroscopic detection of a white dwarf in orbit around the M5 star counterpart of ULPT source GLEAM-X\,J1704$-$37 \citep{2025arXiv250103315R}. The orbital period of 2.9 hrs of this M5/WD system matches the radio pulsation period, suggesting in this case, and also ILT\,J1101+5521 \citep{2024arXiv240811536D}, that the radio properties result from the interaction of close binaries. The radio activity of \hd lacks any obvious periodicity on hour-long timescales, and it is further distinguished from the ULPTs as having bright but steady optical and X-ray counterparts. While we cannot rule out an ULPT origin, there are no strong arguments in its favor either.

% SDH: Do we want to note here, or somewhere, that a period range of 11.5-12.5 days is consistent with the times of the 3 L band detections, assuming an uncertainty of 12 hr in assigning times for each. (I used 21:00 for 2020 June 28 and July 10, and 16:00 for 2024 July 7.) The two MeerKAT nondetections and the three 2024 GMRT L band nondetections do not rule this range out. Well, a 12 day period would be pretty extreme for a ULPT, no??!!

\section{Conclusions and Suggestions for Future Work}\label{sec:conclude}

The radio source \radio was initially identified as an outlier in a search for steep spectrum, polarized point sources in the bulge of our Galaxy. It was found to be positionally coincidence with a bright, nearby K star \hd and nicknamed ``Special K''. Its anomalous radio properties motivated the current study. 

In order to better characterize the radio properties of \hdp, we obtained new GMRT observations, and made use of extensive archival data from ASKAP, MeerKAT and the VLA. Transient radio emission is detected at 1.5 GHz and 3 GHz, but there are significant differences between them. At 1.5 GHz, where there is more extensive monitoring, \hd is active on month-long or week-long timescales, during which there are intervals of several hours duration in which the source is highly variable. The 1.5-GHz emission appears to be periodic, and may be beamed. 
At 3 GHz, short-lived emission is detected only once in five epochs and lasts on timescales of tens of seconds to minutes. The degree of circular polarization and the spectra are very different. At 1.5 GHz the degree of polarization is very large and the spectrum drops sharply above 1 GHz. To our knowledge, this sharp break — reminiscent of spectral breaks seen in magnetized planets of our solar system — is unusual and perhaps unique. Another striking feature of the 1.5-GHz emission from \hd is the apparent constancy of its spectral and polarization properties across different epochs and activity levels. At 3 GHz we place only upper limits on the polarization and the spectrum is relatively flat across the band. Taken together the aggregate radio properties in Table \ref{tab:emprop} argue for two distinct physical emission processes at work; electron cyclotron (ECM) at 1.5 GHz and (possibly) gyro-synchrotron emission at 3 GHz. 

Despite being bright, the K star had been relatively understudied and so we undertook high resolution spectroscopy from CHIRON/SMARTS and speckle interferometric observations from Zorro/Gemini-S. Additionally, we used published optical/NIR photometry, TESS light curves and {\it Gaia} astrometric results. The high resolution optical spectrum was consistent with a mature and chromospherically inactive dwarf star of spectral type K7V. An observed photometric wobble from {\it Gaia} was shown to be consistent with a close-in companion with a period of about 1100 days and a spectral type of either M0.5 or M5.5. Given the {\it Gaia} orbital parameters, we showed that a M0.5 star would have been visible in the diffraction-limited speckle images. As there were no other stars visible in the speckle images we conclude that the secondary was likely a M5.5 star.

We next explored three possible origins for the anomalous radio emission, with special focus on the rare ECM emission. These were (1) Chromospheric activity such as that seen from young, rapidly rotating and/or highly magnetized single stars, or interacting binaries. (2) Auroral (i.e. polar) emission from a star-planet interaction (SPI), or (3) a new class of ultra long-period transients (ULPT). Given their heterogeneous properties, we cannot rule out a ULPT origin, but we find no strong evidence in its favor. Distinguishing between the remaining two models is complicated by not knowing which star is the dominant source of the radio (and X-ray emission). 

The optical spectrum, the absence of optical and X-ray flares, and the rotation period derived from the X-ray/activity relation all support the conclusion that the K star of \hd is a mature, chromospherically inactive dwarf that is unlike any known radio-bright K star. If the ECM emission comes from the K star primary then it must be auroral in origin. However, the preponderance of the evidence favors the M star secondary as the source of the radio emission. This includes the observed radio and X-ray luminosities, the deviation from the {G{\"u}del}--Benz relationship, the appearance of both beamed, periodic ECM and gyro-synchrotron emission. Indeed, one of the difficulties with this rich diversity in phenomenology is in distinguishing between a chromospheric and auroral origin. A M5.5 star lies in the transition region in which both types of emission mechanisms are at work. The absence of any optical (and X-ray) flares gives some support for the auroral interpretation since chromospherically active late-type M stars are known to undergo prodigious flaring. Even if this was shown to be the case, auroral emission is common among the population of radio-emitting UCDs, and it leaves unanswered the larger question of the magnetoplasma engine powering this emission.

An SPI analogous to the Jupiter-Io system could produce auroral emission but it must satisfy at least three criteria (quiescence, periodicity and stability). A convincing SPI case has not been made for any radio stars to date. For \hd the strongest evidence for an SPI lies with the K star, but it is far from proven. The evidence for a quiescent chromosphere is strong and stability of the polarization and spectral index properties point to a stable magnetic field configuration, but the evidence for beamed, periodic emission needs to be confirmed.

In addition to testing whether the ECM emission is beamed and periodic rather than stochastic, there are other future observations that could bolster (or refute) the SPI case. Very Long Baseline Interferometry (VLBI) could
easy resolve the orbit and determine which of the two stars the ECM emission comes from. If the radio emission is shown to originate from the K star then SPI is a compelling hypothesis, while other explanations are possible in the case of the M star. Longer duration TESS observations planned later in 2025 could improve the search for transiting planets and confirm the absence of flares. Alternatively, high resolution spectroscopy could search for a planet around the K star whose period is similar to that of the radio emission. There is also a need to verify the stability of the polarization and spectral properties seen at 1.5 GHz, to ensure that this behavior is not a coincidence but rather is a fundamental property of the emission. Finally, we note that the origin of the 3-GHz emission needs to be better characterized to verify that it is gyro-synchrotron, and a deep search should be made for quiescent radio emission. The X-rays indicate the presence of some coronal emission, which if it follows the {G{\"u}del}--Benz relationship would be about 5 $\mu$Jy. Ultimately, \hd represents a  promising laboratory for studying the interplay of magnetic activity, binarity, and star-planet interactions, and hopefully 
future multi-wavelength observations will uncover new insights into the mechanisms driving its extraordinary radio properties.

\begin{acknowledgments}

We thank Tim Bastian, Gregg Hallinan, Jackie Villadsen and Stephen White, for useful discussions. This research has made use of the SIMBAD database, operated at CDS, Strasbourg, France. The MeerKAT telescope is operated by the South African Radio Astronomy Observatory, which is a facility of the National Research Foundation, an agency of the Department of Science and Innovation. We thank the staff of the GMRT that made these observations possible. GMRT is run by the National Centre for Radio Astrophysics of the Tata Institute of Fundamental Research. The National Radio Astronomy Observatory is a facility of the National Science Foundation operated under cooperative agreement by Associated Universities, Inc. This work has made use of data from the European Space Agency (ESA) mission {\it Gaia} (\url{https://www.cosmos.esa.int/gaia}), processed by the {\it Gaia} Data Processing and Analysis Consortium (DPAC,
\url{https://www.cosmos.esa.int/web/gaia/dpac/consortium}). Funding for the DPAC has been provided by national institutions, in particular the institutions participating in the {\it Gaia} Multilateral Agreement. The material is based in part upon work supported by NASA under award number 80GSFC24M0006. Basic research at NRL is funded by 6.1 Base programs. Construction and installation of VLITE was supported by the NRL Sustainment Restoration and Maintenance fund. This project was supported in part by an appointment to the NRC Research Associateship Program at the US Naval Research Laboratory, administered by the Fellowships Office of the National Academies of Sciences, Engineering, and Medicine. Some of the observations in this paper made use of the High-Resolution Imaging instrument ‘Alopeke and were obtained under Gemini LLP Proposal Number: GN/S-2021A-LP-105. ‘Alopeke was funded by the NASA Exoplanet Exploration Program and built at the NASA Ames Research Center by Steve B. Howell, Nic Scott, Elliott P. Horch, and Emmett Quigley. The TESS data presented in this paper were obtained from the Mikulski Archive for Space Telescopes (MAST) at the Space Telescope Science Institute. The specific observations analyzed can be accessed via \dataset[https://doi.org/10.17909/hzq3-6s3]{https://doi.org/10.17909/hzq3-6s38} and \dataset[https://doi.org/10.17909/f2t0-d696]{https://doi.org/10.17909/f2t0-d696}. STScI is operated by the Association of Universities for Research in Astronomy, Inc., under NASA contract NAS5–26555. Support to MAST for these data is provided by the NASA Office of Space Science via grant NAG5–7584 and by other grants and contracts. This work uses ASKAP data obtained from Inyarrimanha Ilgari Bundara, the CSIRO Murchison Radio-astronomy Observatory. We acknowledge the Wajarri Yamaji People as the Traditional Owners and native title holders of the Observatory site. CSIRO’s ASKAP radio telescope is part of the Australia Telescope National Facility. Operation of ASKAP is funded by the Australian Government with support from the National Collaborative Research Infrastructure Strategy.

%Alopeke was mounted on the Gemini North telescope of the international Gemini Observatory, a program of NSF’s OIR Lab, which is managed by the Association of Universities for Research in Astronomy (AURA) under a cooperative agreement with the National Science Foundation. on behalf of the Gemini partnership: the National Science Foundation (United States), National Research Council (Canada), Agencia Nacional de Investigación y Desarrollo (Chile), Ministerio de Ciencia, Tecnología e Innovación (Argentina), Ministério da Ciência, Tecnologia, Inovações e Comunicações (Brazil), and Korea Astronomy and Space Science Institute (Republic of Korea).

%RS and DK are supported by NSF grant AST-1816904. 

\end{acknowledgments}

\bibliography{specialk_astroph}{}
\bibliographystyle{aasjournal}



\end{document}